\documentclass[aps,prb,twocolumn,floatfix,floats,showpacs,superscriptaddress,raggedbottom]{revtex4}
\usepackage{graphicx}
\usepackage{dcolumn}
\usepackage{amsmath}
\usepackage{url}
\newcommand{\marked}[1]{\textbf{\textsl{#1}}}
\pdfoutput=1

\begin{document}

\title{Coherent switching by detuning a side-coupled quantum-dot system}

\author{Omar Valsson}
\email{omarv@raunvis.hi.is}
\affiliation{Science Institute, University of Iceland, Dunhaga 3,
        IS-107 Reykjavik, Iceland}
\author{Chi-Shung Tang}
\email{cstang@nuu.edu.tw} \affiliation{Department of Mechanical
Engineering, National United University, 1, Lienda, Miaoli 36003,
Taiwan}
\author{Vidar Gudmundsson}
\email{vidar@raunvis.hi.is}
\affiliation{Science Institute, University of Iceland, Dunhaga 3,
        IS-107 Reykjavik, Iceland}

\begin{abstract}
We investigate phase coherent electronic transport in an open
quantum system, which consists of quantum dots side-coupled to a
nanowire. It is demonstrated that coherent switching can be
characterized by adjusting the electronic energy.  A comparative
analysis of quantum coherence effects in side-coupled quantum-dot
systems is presented. Our results demonstrate the relevance of
electronic nanodevices based on coherent switching by
appropriately detuning the side-coupled quantum dots that are
located at variable distance along a wire.
\end{abstract}

\pacs{73.23.-b, 73.21.Hb, 73.21.La, 85.35.Ds}
%

\maketitle

\section{Introduction}
Transport through mesoscopic systems has received enormous interest
within the last decade.\cite{Datta1995,Imry2001}  The advances in
current progress in microfabrication technology have enabled studies
of transport through quantum systems in which the charge carriers
behave coherently.  In recent years, much effort has been devoted to
coherent transport in various coupled quantum-dot and quantum-wire
associated structures such as open quantum
dot,\cite{Bird1999,Moura2002,Tang2003,Mendoza2006,Tang2005} embedded
dot,\cite{Olendski2003,Thorgilsson2007,Muller2007}
side-coupled-dot,\cite{Kang2001,Franco2003,Kobayashi2004,Sato2005,Lee2008}
antidot,\cite{Kirczenow1994,Zozoulenko1996,Merlo2007}
multidot,\cite{Gurvitz1998,Park1999} and coupled quantum-wire
systems.\cite{Chang2003,Tang2006,Huang2008} Rich quantum
interference phenomena in mesoscopic systems have been discussed in
various aspects such as bound-state
features,\cite{Tang1996,Bardarson2004,Gudmundsson05:153306,Ordonez2006,Tang1999}
spin-related switching,\cite{Frustaglia2001,Frustaglia2004,Cui2008}
phase switching,\cite{Sigrist2007,Puller2008} and Andreev current
switching.\cite{Peng2005} Over the last few years, the search for the
quantum transport through quantum-dot systems has made great
progress.  For example, coherent probing experiment for
investigating phase switching of Aharonov-Bohm oscillation in
differential conductance in a two-dot embedded quantum ring has been
performed.\cite{Sigrist2007}   The study of switching effects has
thus drawn a great deal of interest in application and at the
fundamental level.

Theoretically, transport investigations in quantum-dot systems
are often based on the Anderson model.\cite{Anderson1961}
Commonly, it has been assumed that the geometry
is discrete in real space and the coupling
between the system and the external electrodes is relatively
weak.\cite{Ma2000,Costamagna2008,Weymann2008}  The open systems
considered were usually idealized to be strictly one-dimensional
implying the assumption of a single-mode propagation, namely that the
finite width effect of the system was not included.  Alternatively,
the Lippmann-Schwinger formalism assumes a continuous model that has
been used to calculate the electronic and atomic dynamics of
nanoscale conductors under steady-state current
flow.\cite{cattapan:903,Ventra2002} This theory was subsequently
applied considering electronic transport through laterally parallel
double open quantum dots embedded inside a quantum
wire.\cite{Tang2005}  It allowed us to look at arbitrary multidot
potential profiles and illustrate the results by performing
computational simulations for the electronic conductance through the
system with multi-mode propagation.  We found that the coupling
modes of the dots are tunable by adjusting the strength of a central
barrier in the wire.

In this present work, we utilize the Lippmann-Schwinger formalism to
achieve a full quantum mechanical description of the dynamics
of side-coupled quantum-dot systems. We solve a set of coupled
Lippmann-Schwinger equations to express each individual continuum
wave function of the entire system in terms of the corresponding
asymptotic wave function of the source-drain leads, and the
scattering potential representing the mesoscopic system via the
appropriate scattering Green function. Transport regimes arising
from the strong interplay between the dot structures and the open
mesoscopic system are investigated.

The paper is organized as follows. In Sec.\ II we present the
general scheme and description of quantum transport calculation
based on the framework of Lippmann-Schwinger formalism.  In Sec.\
III, the manipulation of the side-coupled quantum dots is addressed.
Coherent switching effects by either tuning the electronic energy or
detuning one of the side-coupled dots are discussed.   Finally, in
Sec.\ IV we shall summarize and draw conclusions.

\section{Lippmann-Schwinger Model}
\begin{figure}[tbhq]
 \centering
 \includegraphics[width=0.45 \textwidth]{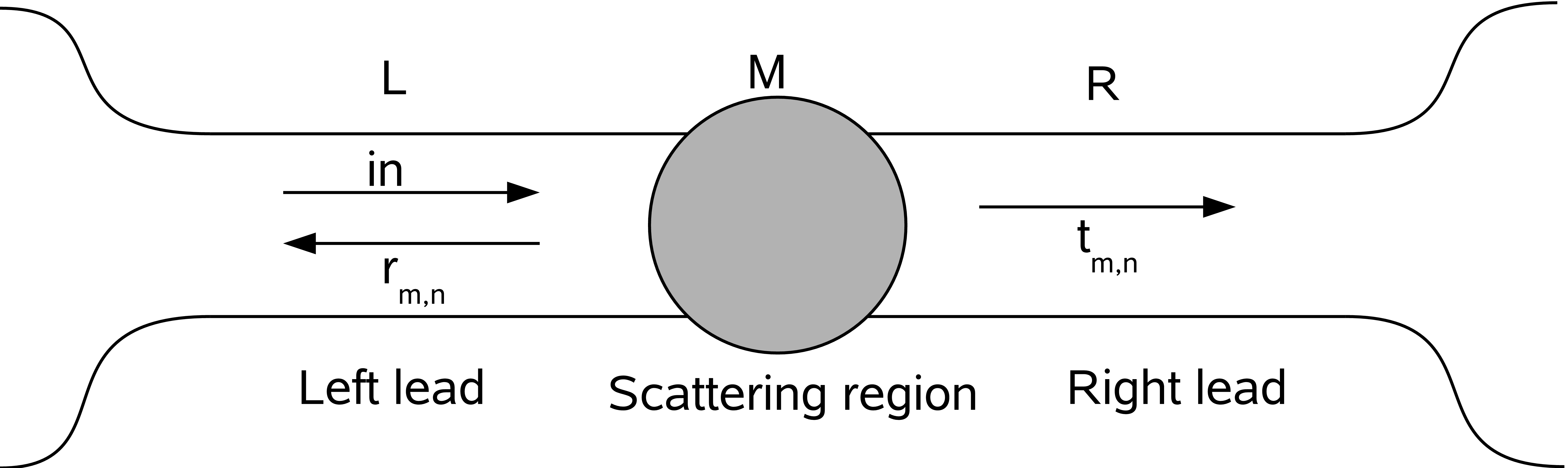}
 \caption{Schematic diagram of a scattering region in a nanowire
          connected to electronic reservoirs.}
 \label{Fig:qwire_setup}
\end{figure}

The physical system under investigation is a two-dimensional
nanowire with side-coupled quantum dots. The considered infinitely
long wire allows electrons to propagate along the wire direction while being
confined transversely. The considered mesoscopic system can be
viewed as three divided regions, namely the left source lead, the
scattering region including the side-coupled dots, and the right drain
lead, as is depicted in Fig.~\ref{Fig:qwire_setup}.
The Hamiltonian describing the system can be written as
\begin{equation}
H = H_0 + V_{\mathrm{sc}}(x,y),
\end{equation}
where $ V_{\mathrm{sc}}$ is the scattering potential
describing the side-coupled quantum dots,
localized inside the scattering region.
The unperturbed Hamiltonian is given by
\begin{equation}
 \label{Eq:H0}
H_0 =   -\frac{\hbar^2}{2m^{*}} \nabla^2
 + V_{\mathrm{c}}(y),
\end{equation}
where $m^{*}$ stands for the effective mass of the electron.  The
confining potential $V_{\mathrm{c}}(y)$ is assumed of the parabolic
form
\begin{equation}
 V_{\mathrm{c}}(y) = \frac{1}{2} m^{*} \omega^{2} y^{2},
\end{equation}
leading to the discrete subband energy levels
\begin{equation}
\varepsilon_{n} = \left(n + \frac{1}{2}\right) \hbar \omega ,
\end{equation}
with $n$ = $0,1,2,\dotsc$ and level spacing $\Delta \varepsilon_n$ =
$\hbar \omega$.

For the $n$-th mode and without magnetic field, the unperturbed
eigenfunctions of $H_0$ can be expressed as  $\exp[{\pm i k_{n} x}] \chi_n
(y)$, where $\chi_n (y)$ stands for the confinement modes of the
nanowire, determined by
\begin{equation}
\label{Eq:SubbandWavefunctions}
 \biggl[
 - \frac{\hbar}{2 m^{*}} \frac{d^2}{d y^2} + V_{\mathrm{c}}(y)
 \biggr]
 \chi_{n}(y) = \varepsilon_{n} \chi_{n}(y),
\end{equation}
and the wave number $k_{n}$ is given by
\begin{equation}
\label{Eq:WaveNumberMode}
 k_{n} =
 \sqrt{
   \frac{2 m^{*}}{\hbar^2}
   \big( E-\varepsilon_{n} \big)
      }.
\end{equation}
We would like to mention that the real wave number $k_{n}$ implies
the $n$-th mode being a propagating mode, otherwise it indicates an
evanescent mode that plays an important role to quantum interference
in coherent transport.  For an electron incident from the source
electrode with energy $E$ occupying the mode $n$, the electron
may be reflected into mode $m$ with amplitude $r_{m,n}$ or
transmitted into mode $m$ with amplitude $t_{m,n}$ to the drain
electrode. The corresponding wave function describing the scattering
processes can be expressed as
\begin{eqnarray}
 \Psi_{E,n}(x,y) &=&
  \exp\left({ i k_{n} x}\right)
  \chi_n (y)
  \nonumber \\
  &&+
  \sum\limits_{m} r_{m,n}
  \exp\left(-i k_{m} x\right)
  \chi_m (y),
\end{eqnarray}
if $(x,y) \in L$;
\begin{equation}
 \Psi_{E,n}(x,y) =
  \sum\limits_{m} \varphi_{E,n;m}(x) \chi_{m}(y),
\end{equation}
if $(x,y) \in M$; and
\begin{equation}
 \Psi_{E,n}(x,y) =
  \sum\limits_{m} t_{m,n}
  \exp\left(i k_{m} x\right)
  \chi_m (y),
\end{equation}
if $(x,y) \in R$.  Here $L$, $M$, and $R$ denote the left source lead, the
scattering region, and the right drain lead, respectively.  In addition,
$\varphi_{E,n;m}(x)$ represents the $x$ component of the $m$-th mode
scattering wave function in the scattering region.
This allows us to expand the Schr\"odinger equation in the
scattering region into a set of coupled differential equations
\begin{equation}
 \label{Eq:Coupled_ODE}
 \bigg(
 \frac{d^2}{d x^2} + k_{m}^2
 \bigg) \varphi_{E,n;m}(x)
 =
 \frac{2 m^{*}}{\hbar^2}
 \sum_{m'} V_{m,m'}(x) \varphi_{E,n;m'}(x),
\end{equation}
in which all evanescent and propagating intermediate modes $m'$ are taken
into account.  The coupling between these intermediate modes is
characterized by the matrix elements of the scattering potential,
namely
\begin{equation}
 \label{Eq:MatrixElements}
 V_{m,m'}(x) =
 \int d y \, \chi_{m}^{*}(y) V_{\mathrm{sc}}(x,y) \chi_{m'}(y),
\end{equation}
Here, the matrix elements are calculated analytically.

To obtain the Lippmann-Schwinger equation for
handling the transport properties,  we start from the unperturbed
Green function, that obeys the differential equation
\begin{equation}
 \label{Eq:GreensFunction_ODE}
 \bigg(
 \frac{d^2}{d x^2} + k_m^2
 \bigg)
 G_{E,m}^{0}(x,x') = \delta(x-x'),
\end{equation}
describing the electron incident from the source electrode. This
unperturbed Green function can be expressed as
\begin{equation}
 \label{Eq:GreensFunction}
 G_{E,m}^{0}(x,x') =
 \frac{1}{2i k_{m}}
 \exp\left(i k_{m} \vert x-x' \vert \right),
\end{equation}
and is utilized to transform from the
differential equations (\ref{Eq:Coupled_ODE}) into an infinite set
of coupled Lippmann-Schwinger equations
\begin{widetext}
\begin{equation}
 \label{Eq:Coupled_LippmannSchwinger}
 \varphi_{E,n;m}(x)  =
 \delta_{n,m} \exp\left(i k_{m} x\right) +
 \frac{2 m^{*}}{\hbar^2}
 \sum_{m'} \int d x'
 G_{E,m}^{0}(x,x')
 V_{m,m'}(x') \varphi_{E,n;m'}(x').
\end{equation}
The transmission amplitude $t_{m,n}$ for transmission from the
$n$-th mode to the $m$-th mode can be found by looking at the
asymptotic limit $x \gg 0$, resulting in
\begin{equation}
 \label{Eq:TransmissionAmplitudes}
 t_{m,n}(E)  =
 \delta_{m,n} +
 \frac{m^{*}}{i \hbar^2 k_{m}} \sum_{m'} \int dx'
 \exp\left(-i k_{m} x'\right)
 V_{m,m'}(x') \varphi_{E,n;m'}(x') ,
\end{equation}
where $m$ and $n$ are propagating modes.
\end{widetext}

Based on the framework of the well known Landauer-B\"uttiker
formalism,\cite{Landauer1957,Buttiker1983} the transmission
probability
\begin{equation}
 T_{m,n} =  \frac{k_{m}}{k_{n}}  \vert t_{m,n}(E) \vert^2
\end{equation}
can be calculated numerically at the Fermi surface, and the
conductance is related to the transmission,\cite{Imry1999} expressed
as
\begin{equation}
 \label{Eq:LandauerButtiker}
 G(E) =  \frac{2e^2}{h} \sum_{n,m}  T_{m,n}(E),
\end{equation}
where the sum is over all propagating modes.

It is worth mentioning that the solution of the coupled
Lippmann-Schwinger equations (\ref{Eq:Coupled_LippmannSchwinger})
is achieved by putting the $x$ variable on a grid.  The
integrals are computed by using extended Bode's
rule~\cite{Abramowitz_1965} for numerical integration, taking care
of the cusp in the Green function (\ref{Eq:GreensFunction}) at
$x=x'$ before integrating.

Finally, we note that the above
Lippmann-Schwinger model is quite general
towards the choice of a confinement and scattering potentials.

\section{Results and discussion}
To explore the electronic transport properties in an open mesoscopic
system consisting of a nanowire influenced by side-coupled quantum
dots, we assume the system to be fabricated in a high-mobility
GaAs-Al$_x$Ga$_{1-x}$As heterostructure forming a two-dimensional
electron gas such that the effective mass of an electron is
$m^{*}=0.067 \, m_0$ with $m_0$ being the free electron mass. In
addition, we select the confinement parameter such that the energy
level spacing is $\Delta\varepsilon_n=\hbar \omega=1$~meV.  In the
numerical calculation, it is convenient to redefine the energy scale
by the lowest subband energy $\varepsilon_0 = \hbar
\omega/2=0.5$~meV and the length scale by the characteristic length
$a_{\omega}  = \sqrt{m^{*}\omega/\hbar}= 33.7$~nm.  In all the
figures shown below, we have defined the energy-related parameter
\begin{equation}
X_{E} = \frac{E}{2\varepsilon_0} + \frac{1}{2} ,
\end{equation}
where the integral part of $X_{E}$ indicates the number of
propagating modes at the energy $E$.\cite{Tang1996}
We note that numerical accuracy was always carefully checked
and the calculations were performed using in total
20 subbands and 301 grid points. In the
following, we shall investigate the dynamic motion of the electronic
wave in a quantum wire with side-coupled quantum dots.

\subsection{Single and double side-coupled quantum dot systems}

A side-coupled quantum dot system can be regarded as an open
mesoscopic system with transport characteristics that are tunable by
a kind of side-stub nanostructures.  In contrast to the embedded
quantum-dot system, the transmission through the side-coupled
quantum-dot system consists of a quantum interference between the
ballistic nanowire and the resonant channels in the side-coupled
nanostructures.  Recent works dealing with side-coupled
quantum-dot system rely mainly on an Anderson model in which
the side-coupled dots are assumed to behave effectively like
artificial quantum impurities.\cite{Malyshev2006,checkZitko2006}  It
is thus warranted to devote further effort in developing numerical
techniques in order to analyze the quantum behavior of the
electronic transport in a side-coupled quantum-dot system with an
expected smooth profile.

\begin{figure}[htbq]
 \centering
 \includegraphics[width=0.45 \textwidth]{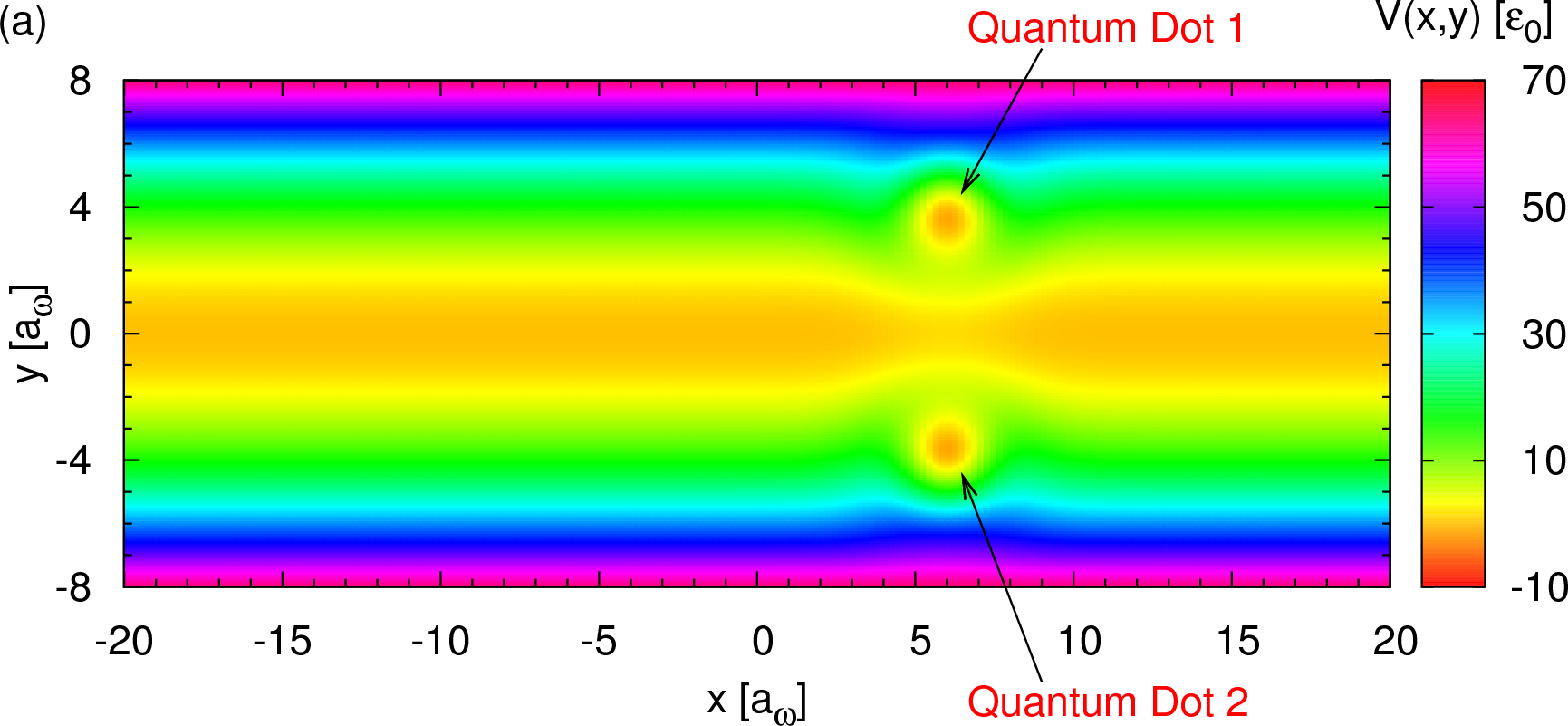}
 \includegraphics[width=0.45 \textwidth]{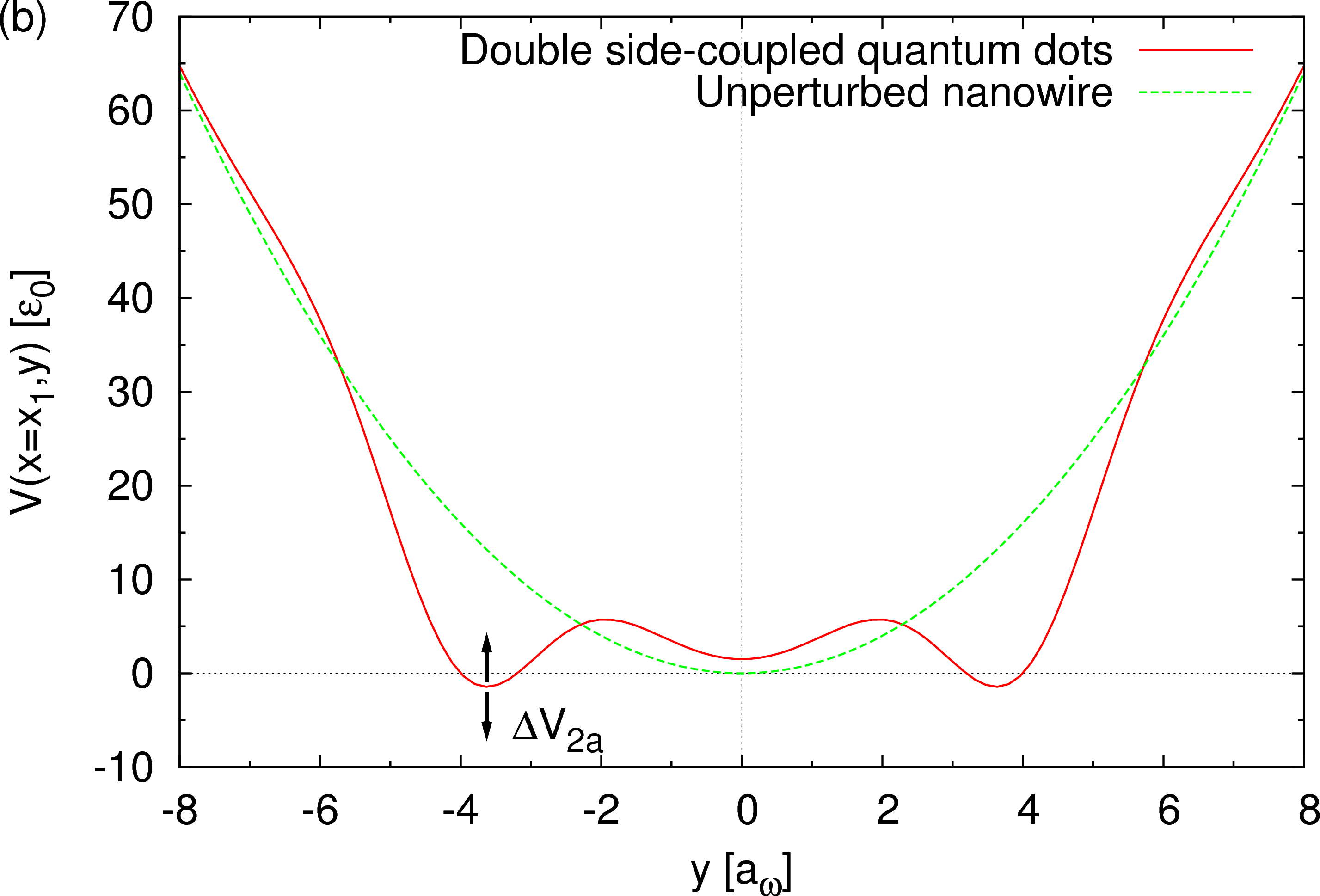}
 \caption{(Color online) Diagram of the system under
 investigation. (a) Double quantum dots side-coupled to a nanowire.
 (b) Transverse potential profile at $x=x_1$ for the double quantum dots
 side-coupled to a nanowire (solid red) in comparison with the unperturbed
 nanowire (dashed green).}
 \label{Fig:TwoP_SCQD_Potential}
\end{figure}

We start by modeling a quantum wire with single or double
side-coupled quantum dots, placed symmetrically in parallel on each
side of the nanowire. The scattering potential describing the pair
of quantum dots is modeled by a linear combination of four Gaussian
functions
\begin{eqnarray}
 V_{\mathrm{sc}}(x,y) & = &
 V_{1a} e^{-[\beta_{x,1a}(x-x_{1})^2+\beta_{y,1a}(y-y_{1})^2]}
 \nonumber\\
 &&+
 V_{1b} e^{-[\beta_{x,1b}(x-x_{1})^2+\beta_{y,1b}(y-y_{1})^2]}
 \nonumber\\
 &&+
 V_{2a} e^{-[\beta_{x,2a}(x-x_{2})^2+\beta_{y,2a}(y-y_{2})^2]}
 \nonumber\\
 &&+
 V_{2b} e^{-[\beta_{x,2b}(x-x_{2})^2+\beta_{y,2b}(y-y_{2})^2]},
\end{eqnarray}
where the first two terms describe the geometry of quantum dot
1 and the last two terms are associated to quantum dot 2, as is
depicted in Fig.\ \ref{Fig:TwoP_SCQD_Potential}(a).  In this
subsection, we focus on two identical side-coupled dots, namely with
$V_{1a} = V_{2a} =-36$ and $V_{1b} = V_{2b} =20$ in units of
$\varepsilon_0$.  In addition, we assume that the two quantum dots
are in parallel and at the same distance on opposite sides of the
nanowire, centered at $(x_1, y_1)= (6, 4)$ and $(x_2, y_2)= (6, -4)$
in units of $a_{\omega}$, respectively.  Moreover, the form factors
describing the broadening of the quantum dots are selected as $
\beta_{x,1a} = \beta_{y,1a} = \beta_{x,2a} = \beta_{y,2a} = 0.4 $
and $\beta_{x,1b} = \beta_{y,1b} = \beta_{x,2b} = \beta_{y,2b} =
0.2$ in units of $a_{\omega}^{-2}$, respectively.  In Fig.\
\ref{Fig:TwoP_SCQD_Potential}(b), we show the transverse profiles
due to the contributions of the confining and the scattering
potentials at $x=x_1$.  The dashed-green curve indicates the
transverse potential profile of the unperturbed nanowire, and the
double side-coupled dot system has a potential profile illustrated
by the solid-red curve.  Later on, we shall adjust the parameter
$V_{2a}$ of the lower quantum dot 2 by $\Delta
V_{2a}$, thus varying the depth of dot 2, for studying the sensitivity and
dynamics of quantum interference effects.

\begin{figure}[htbq]
 \centering
 \includegraphics[width=0.45 \textwidth]{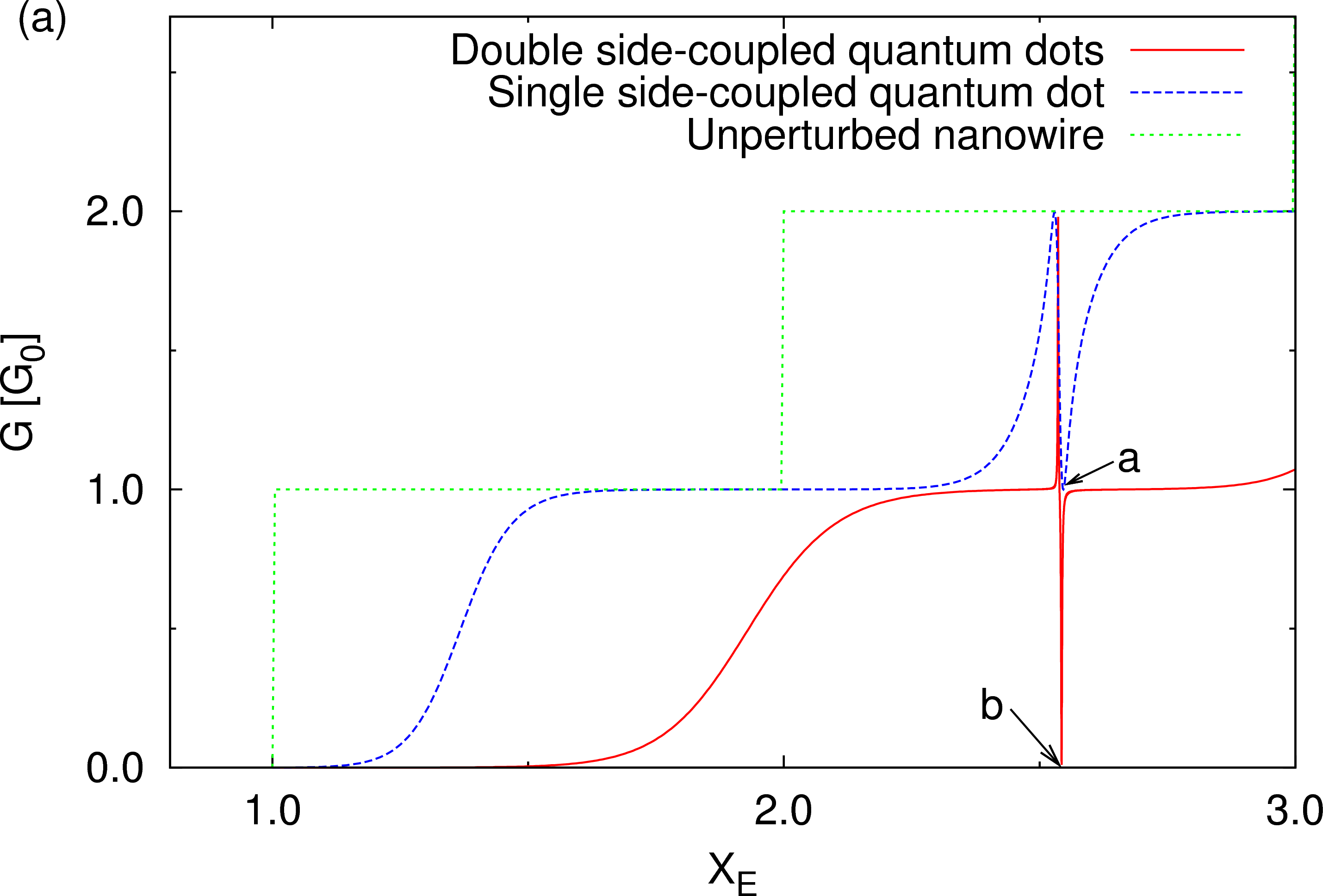}
 \includegraphics[width=0.45 \textwidth]{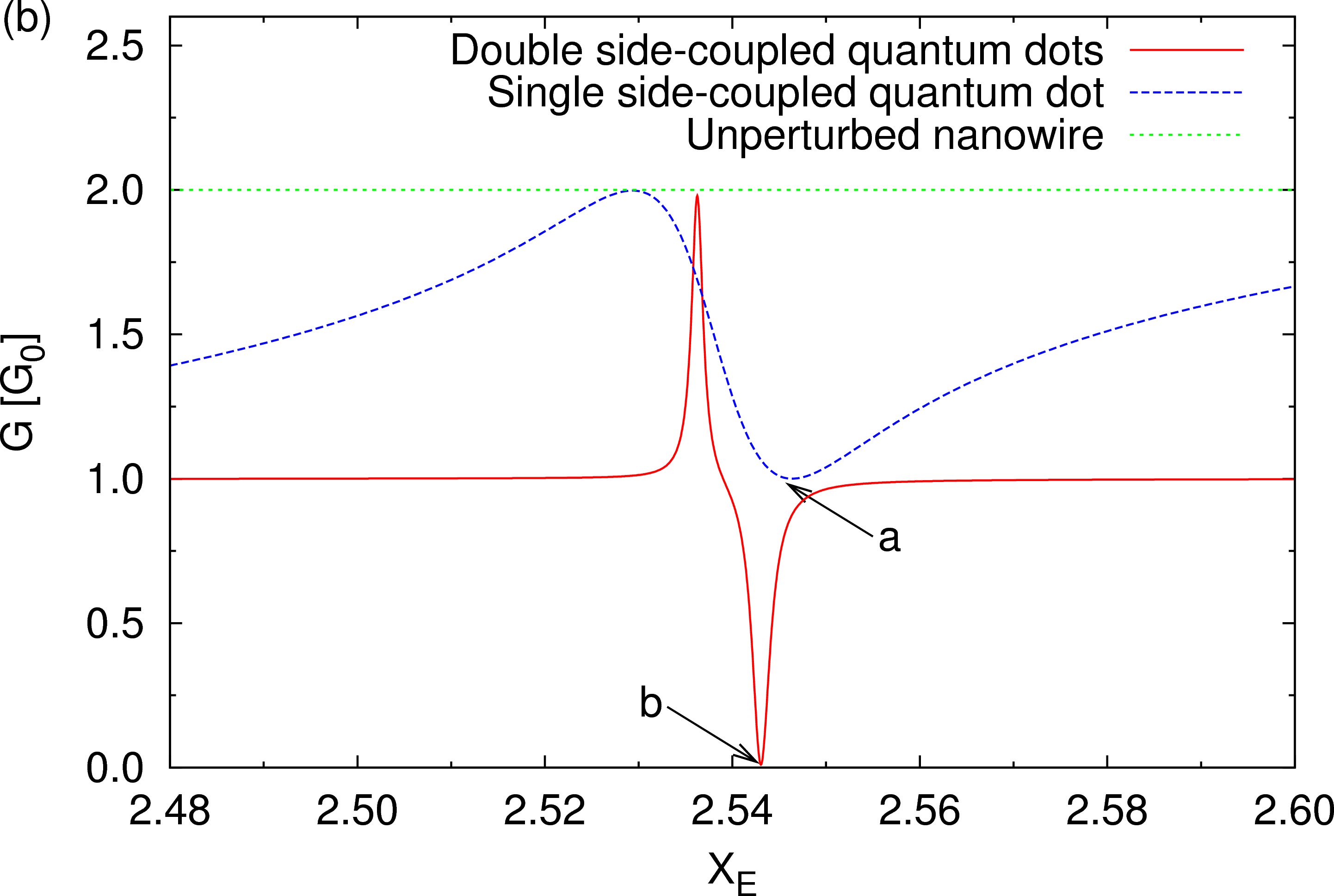}
 \caption{(Color online) (a) Conductance as a function of the
 parameter $X_{E}$ for the cases of double side-coupled quantum dots (solid red),
 single side-coupled quantum dot (dashed blue), and unperturbed nanowire (dotted
 green). The conductance curves exhibiting Fano resonances are emphasized in (b).}
 \label{Fig:TwoP_SCQD_EneData}
\end{figure}
In Fig.~\ref{Fig:TwoP_SCQD_EneData} we show the conductance as a
function of the parameter $X_{E}$ for the lowest two conductance
plateaus for the case of double side-coupled dots (solid red) in
comparison with the case of a single side-coupled dot (dashed blue)
and the unperturbed nanowire (dotted green).  In the absence of a
side-coupled quantum dots, the conductance of the ideal nanowire
reproduces the well known conductance quantization plateaus in units
of $G_0 = 2e^2 /h$.  For the case of a single side-coupled dot and
the instate in the low kinetic energy regime, the conductance is
strongly suppressed due to an enhanced backscattering by the
quantum dot.  The backscattering effect is blue shifted by including
the second quantum dot. This interesting conductance suppression
effect can be applied as a conductance switching at
$X_E \approx 1.0$ by turning on and off the quantum dot 2 by
tuning a gate voltage used to form it.

In the mediate kinetic energy regime of the second subband region,
there is a clear antisymmetric peak-and-dip structure in the conductance
around  $X_{E} \approx 2.54$, exhibiting the well-known Fano
resonance feature.\cite{Fano1961} This is attributed to quantum
interference between the continuum in the wire and the quasibound
state in the side-coupled quantum dots.  Both in the single and
double side-coupled-dot systems, the Fano peaks can reach $2 \,
G_0$. However, the Fano dip of the single-dot case reaches $G_0$ but
the Fano dip of the double-dot case may be suppressed to zero
conductance.  This implies that only the electrons occupying the
second subband contribute to the resonance reflection for the case
of single-dot, however for the case of double-dot the electrons
occupying both subbands may contribute to the resonant reflection
leading to the $2 \, G_0$ drop from the Fano peak.

\begin{figure}[htbq]
 \centering
 \includegraphics[width=0.45 \textwidth]{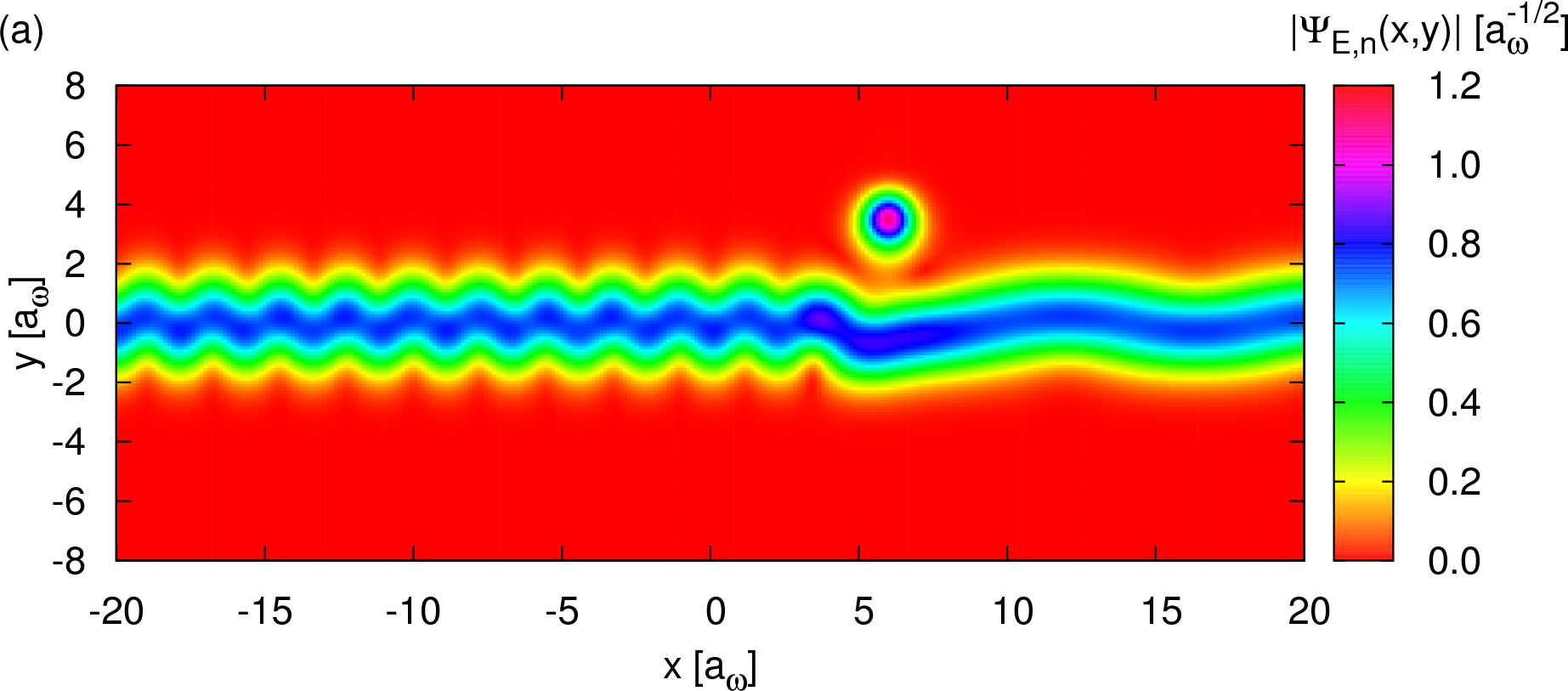}
 \includegraphics[width=0.45 \textwidth]{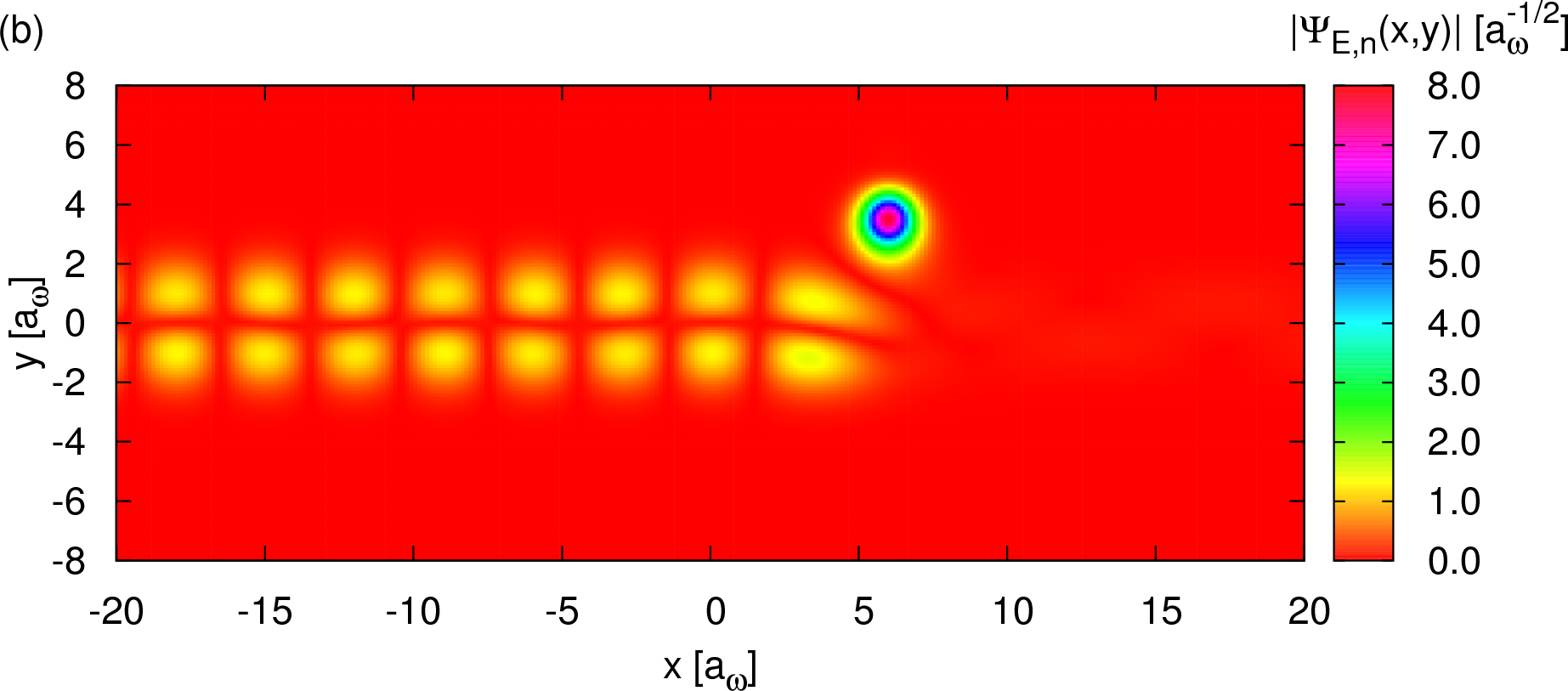}
 \caption{(Color online) Square root of probability densities of the scattering
 states for the case of single-dot system
          corresponding to the Fano resonance dip marked by \marked{a} in
          Fig.~\ref{Fig:TwoP_SCQD_EneData} for the electrons occupying the incident subband (a) $n$=$0$; and (b) $n$=$1$.}
 \label{Fig:Psi_a}
\end{figure}

\begin{figure}[htbq]
 \centering
 \includegraphics[width=0.45 \textwidth]{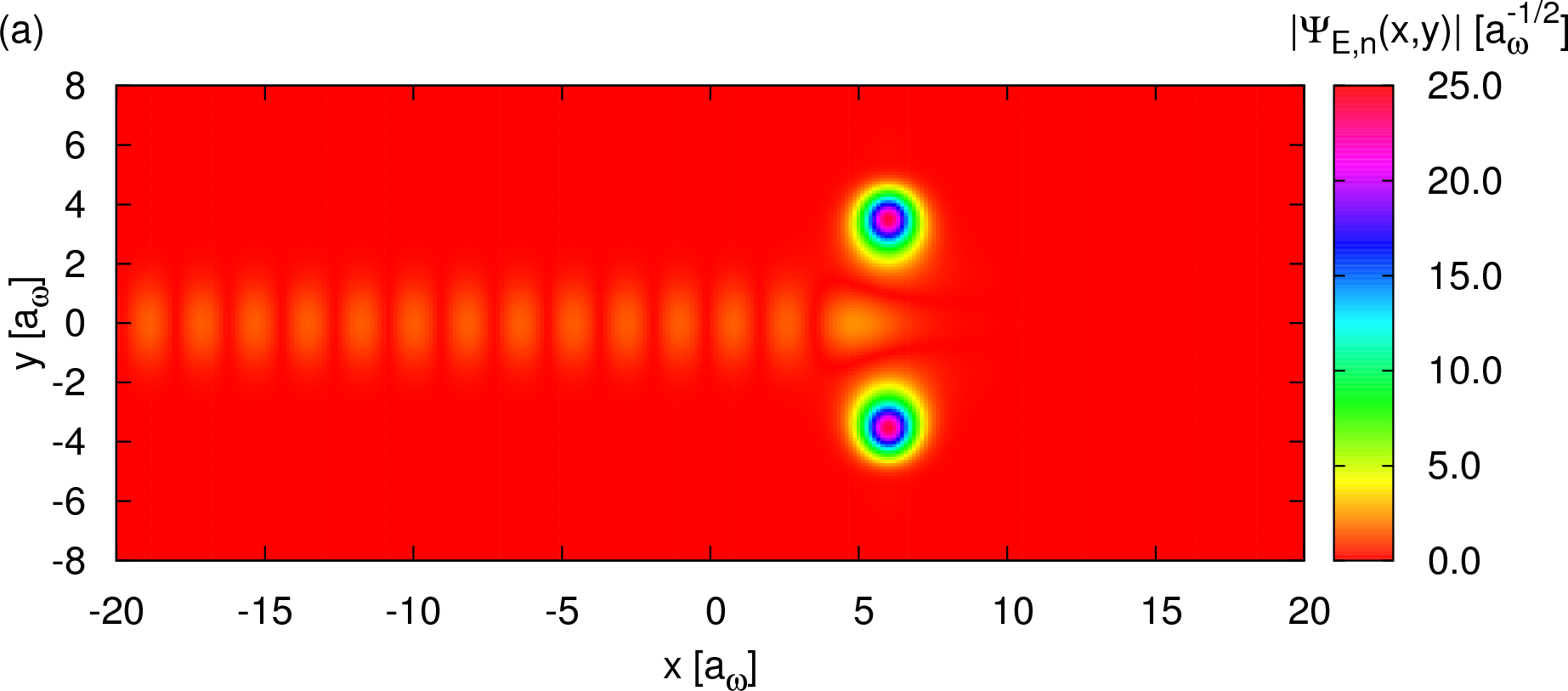}
 \includegraphics[width=0.45 \textwidth]{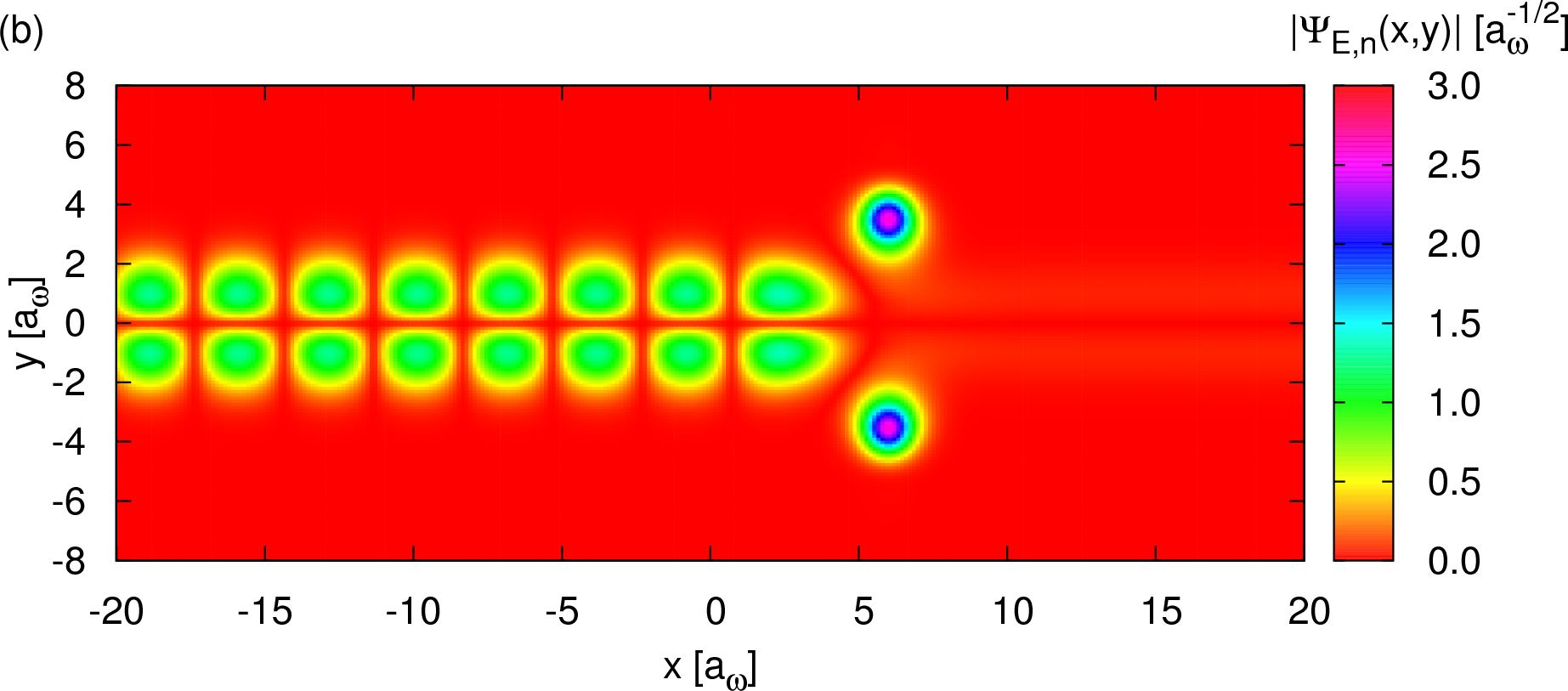}
 \caption{(Color online) Square root of probability densities of the scattering
 states for the case of double-dot system corresponding to the Fano resonance dip marked by \marked{b} in
Fig.~\ref{Fig:TwoP_SCQD_EneData} for the electrons occupying the
incident subband (a) $n$=$0$; and (b) $n$=$1$.}
 \label{Fig:Psi_b}
\end{figure}

To demonstrate the above Fano resonance behavior in conductance we
show, in Figs.~\ref{Fig:Psi_a} and~\ref{Fig:Psi_b}, the square root
of the probability densities of the scattering states at the dip
structures in Fig.~\ref{Fig:TwoP_SCQD_EneData} marked by \marked{a}
($E=4.0927 \, \varepsilon_0$) for the case of a single side-coupled
dot and the dip marked by \marked{b} ($E=4.0861 \, \varepsilon_0$)
for the case of double side-coupled dots, respectively.  For the
single-dot case, the electron occupying the first subband ($n$=$0$)
are in an extended state with negligible coupling to the upper dot,
however the electron occupying the second subband ($n$=$1$) couple
strongly to the upper dot and are resonantly reflected.  For
the case of double side-coupled dots, both the electrons occupying
the first and the second subbands are able to couple strongly to the
side-coupled dots and are resonantly reflected.  Consequently, for the
cases of a single and double side-coupled quantum dots, the
strengths of the Fano lineshapes are $G_0$ or $2 \, G_0$,
respectively. We would like to mention in passing that the
scattering states for the peak and the dip are very similar, showing
a strong and long-lived quasibound state feature that is localized
in each of the two quantum dots. The shape of the quasibound states
indicates that these are the lowest quasibound states in the quantum
dots.

\subsection{Detuning effects in a double side-coupled quantum dot
system}

Double quantum dot systems are good candidates for revealing
coherent quantum transport properties.  Energy detuning between the
different quantum-dot levels can be experimentally investigated by
using charging diagram measurements or using an excitation
spectroscopy technique.\cite{Kouwenhoven1997}  From application
point of view, it was found that the conduction in a cavity mode
with detuning parameter may be utilized as a quantum bit
readout.\cite{Wallraff2004} It is thus warranted to investigate the
detuning effects by considering a parallel side-coupled double-dot
system.

The physical parameters considered here are the same as in the previous
subsection such that the two side-coupled dots are identical and the
system remains spatially symmetric.  The lower quantum dot 2 is
adjusted by the parameter $\Delta V_{2a}$ by tuning a gate voltage
so that the energy levels of the lower quantum dot 2 are relatively
detuned and the spatial symmetry of the system is broken.  As a
result, the symmetry of the quasibound state feature causing the
Fano resonance (see Fig.\ \ref{Fig:TwoP_SCQD_EneData}) tends to be
destroyed by increasing the detuning potential.   Comparing to
overall scattering potentials and characteristic energies, the
detuning parameters are assumed to be relatively small such that the
detuning only has an effect on the resonance structure around $X_{E}
\approx 2.5$.

\begin{figure}[htbq]
 \centering
 \includegraphics[width=0.45 \textwidth]{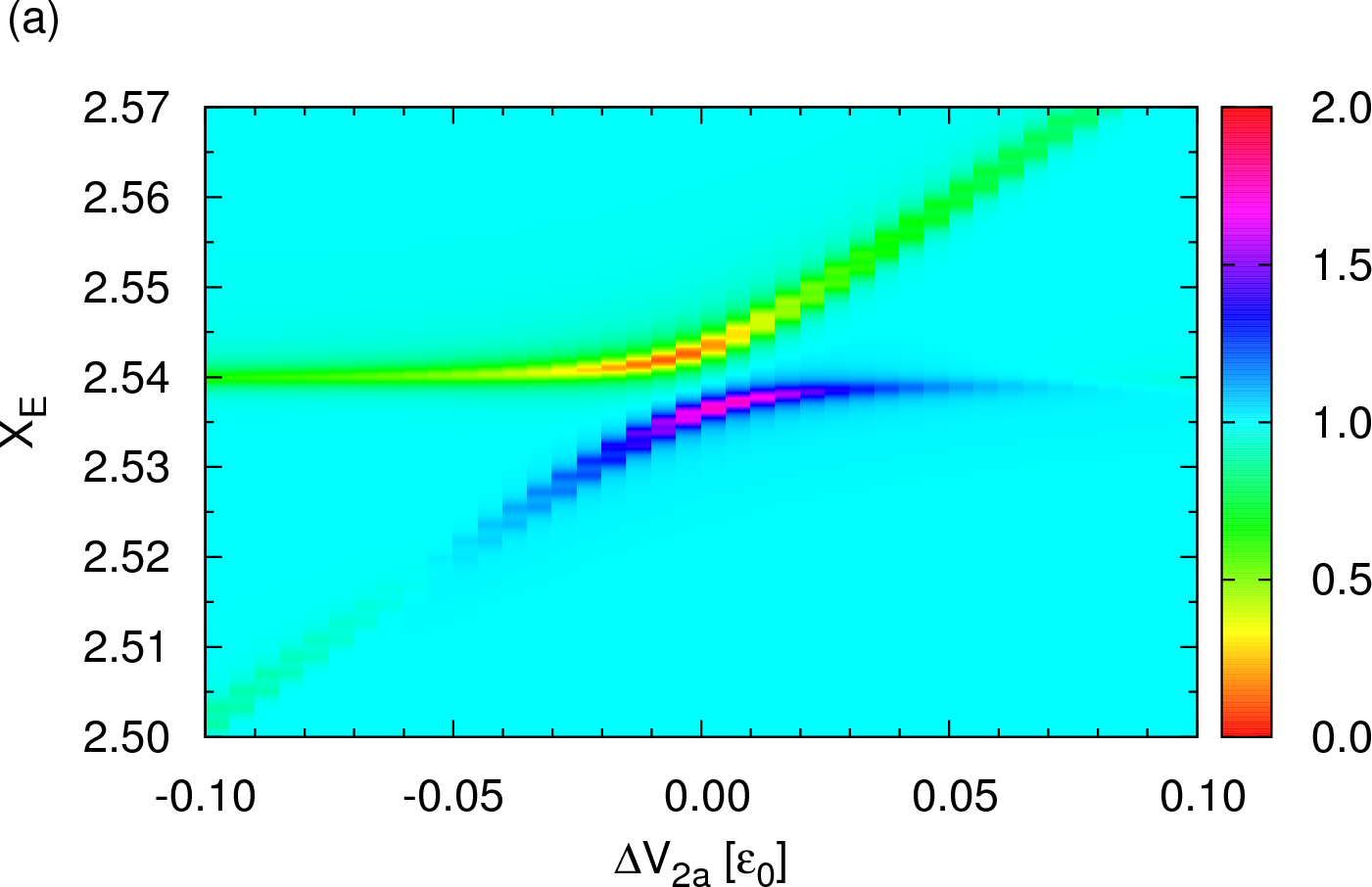}
 \includegraphics[width=0.45 \textwidth]{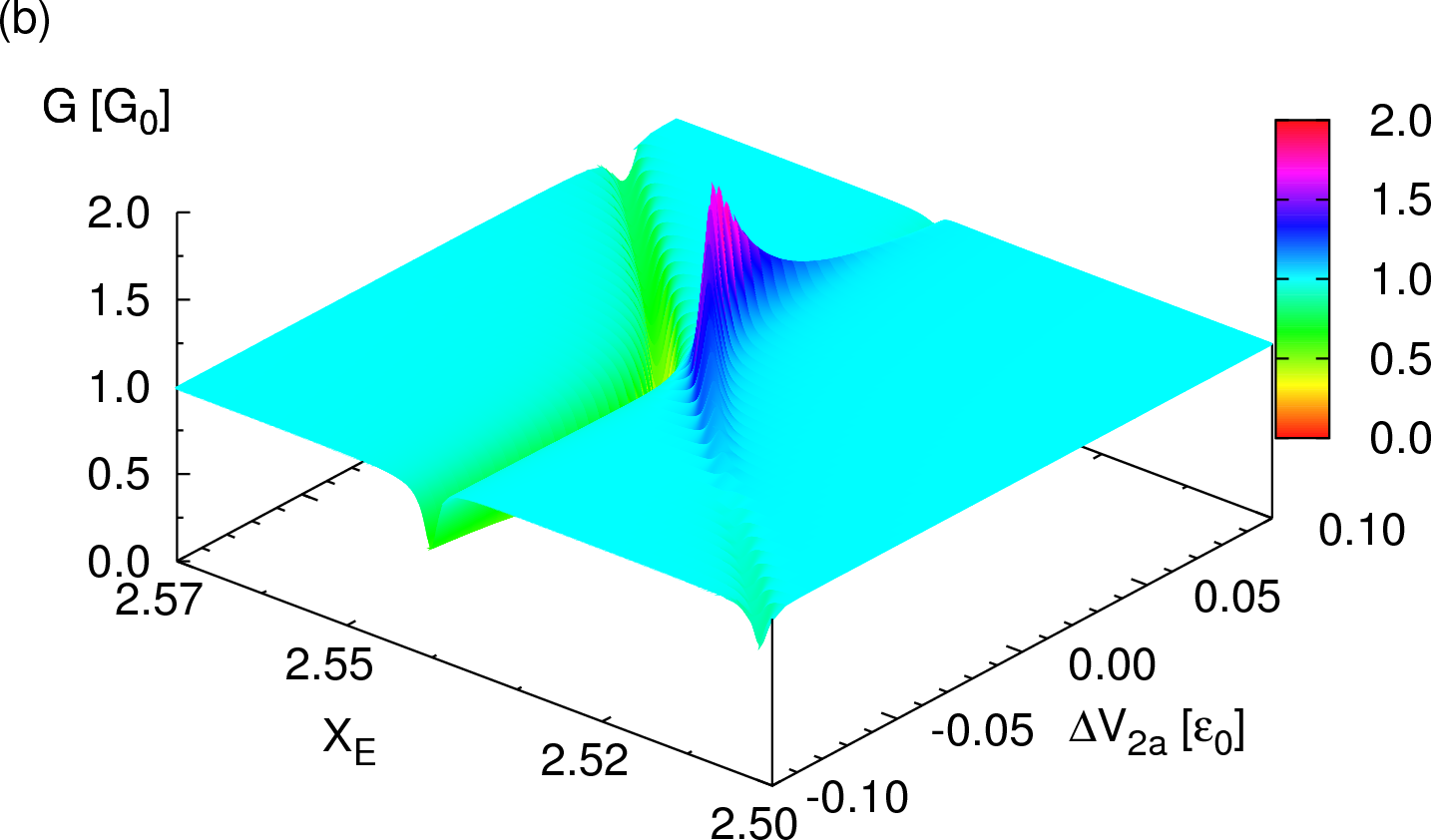}
 \caption{(Color online) Conductance as functions of the
          parameter $X_{E}$ and detuning parameter $\Delta V_{2a}$
          for the case of double side-coupled quantum dots:
          (a) Top view; (b) 3D view.}
 \label{Fig:TwoP_SCQD_EneData_DT1}
\end{figure}

Figure~\ref{Fig:TwoP_SCQD_EneData_DT1} demonstrates the conductance
as a function of the parameter $X_{E}$ as well as the detuning
parameter $\Delta V_{2a}$.  The deviation of the peak and the dip
structures forming the Fano lineshape tends to be suppressed by
increasing the detuning potential.   Both the peak and the dip
structures get less pronounced and the peak may crossover to a dip
when the detuning parameter $\Delta V_{2a}$ is sufficiently large.
Also, we observe an avoided crossing of the peak and the dip during
variation of the detuning parameter through zero.

\begin{figure}[htbq]
 \centering
 \includegraphics[width=0.45 \textwidth]{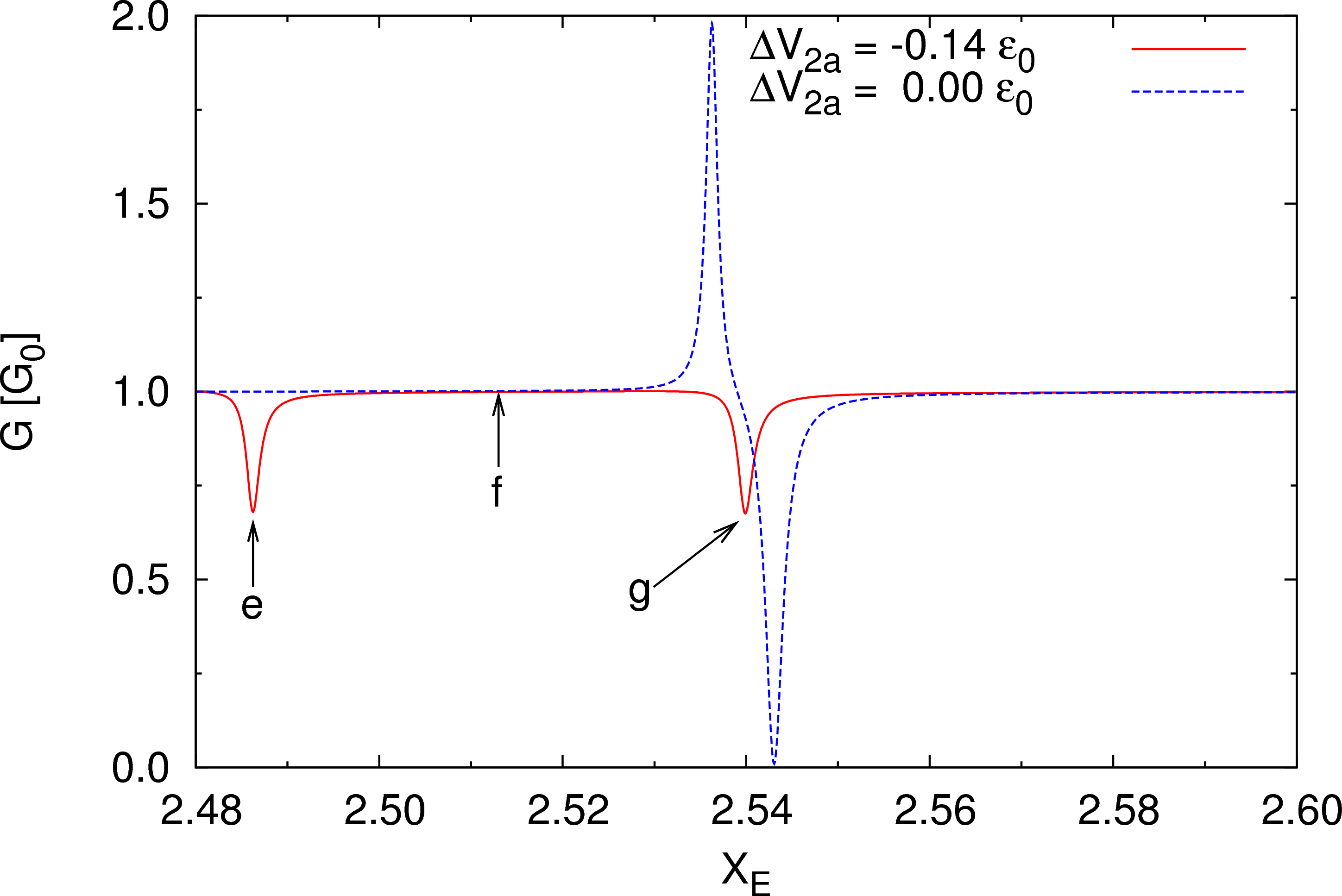}
 \caption{(Color online) Conductance as a function of the
          parameter $X_{E}$ of the double side-coupled
          quantum dots for the cases with detuning
          $\Delta V_{2a} = -0.14 \, \varepsilon_0$ (solid red)
and with no detuning $\Delta V_{2a} = 0.0 \, \varepsilon_0$ (dashed
blue).}
 \label{Fig:TwoP_SCQD_EneData_DT2}
\end{figure}

\begin{figure}[htbq]
 \centering
 \includegraphics[width=0.45 \textwidth]{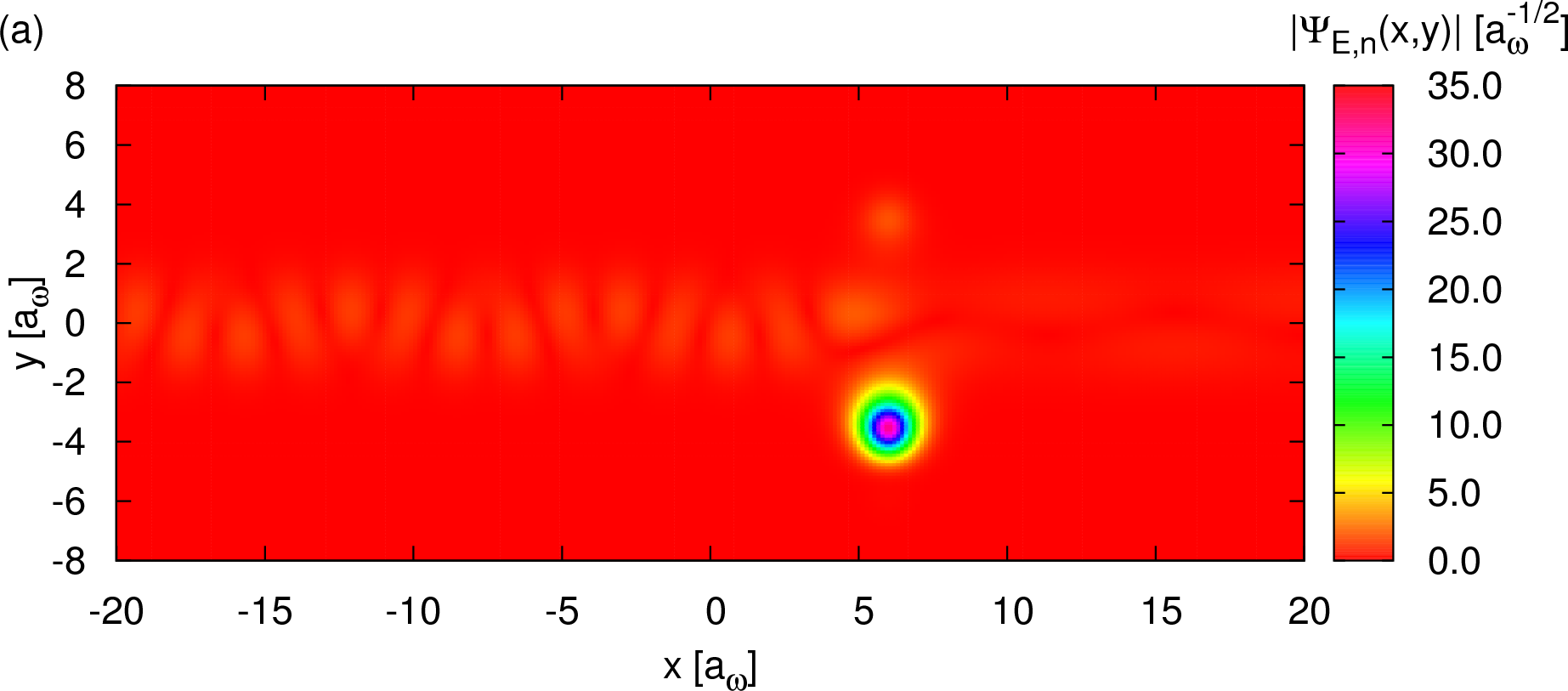}
 \includegraphics[width=0.45 \textwidth]{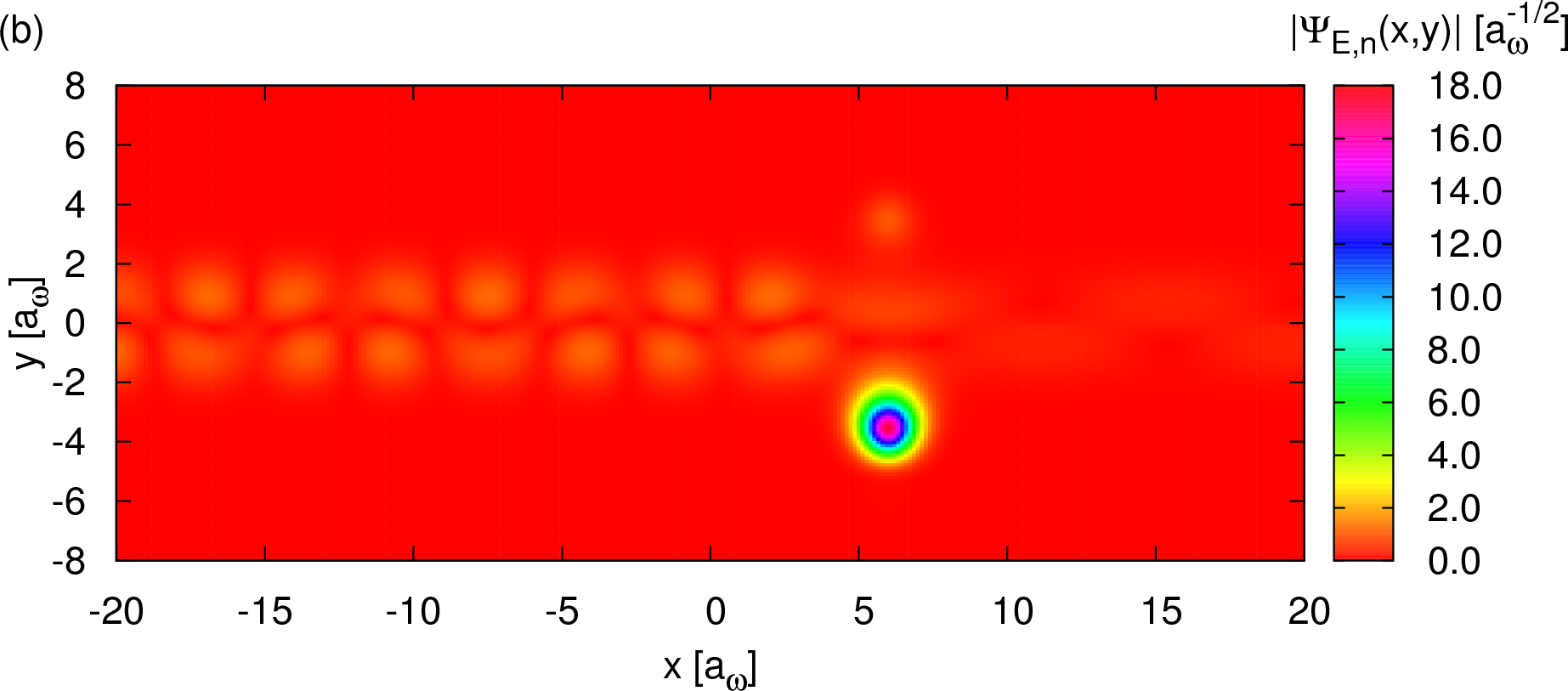}
 \caption{(Color online) Square root of probability densities of the scattering states
          at the dip structure in conductance marked by \marked{e} in
          Fig.~\ref{Fig:TwoP_SCQD_EneData_DT2} for the case of $\Delta V_{2a} = -0.14 \,
          \varepsilon_0$ with incident subband (a) $n$=$0$; and (b) $n$=$1$.}
 \label{Fig:Psi_e}
\end{figure}

\begin{figure}[htbq]
 \centering
 \includegraphics[width=0.45 \textwidth]{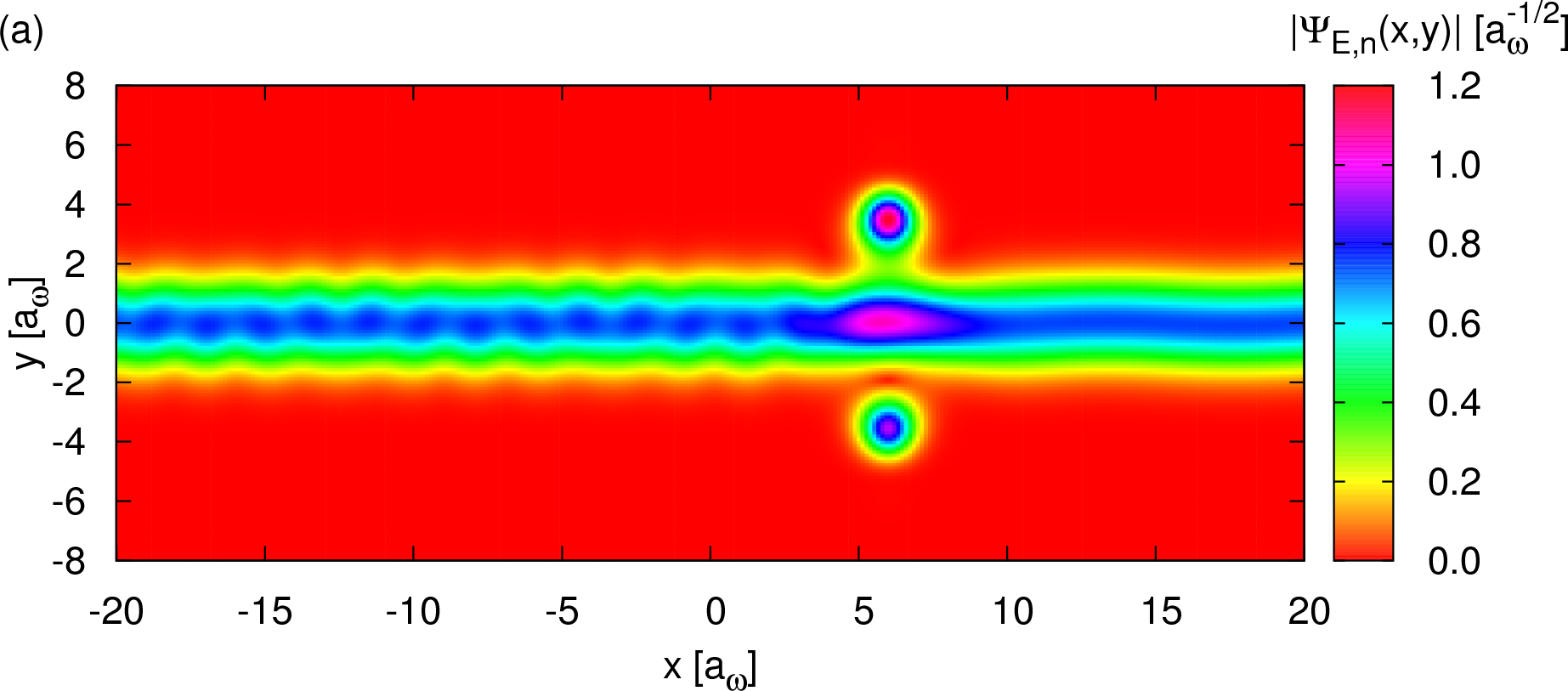}
 \includegraphics[width=0.45 \textwidth]{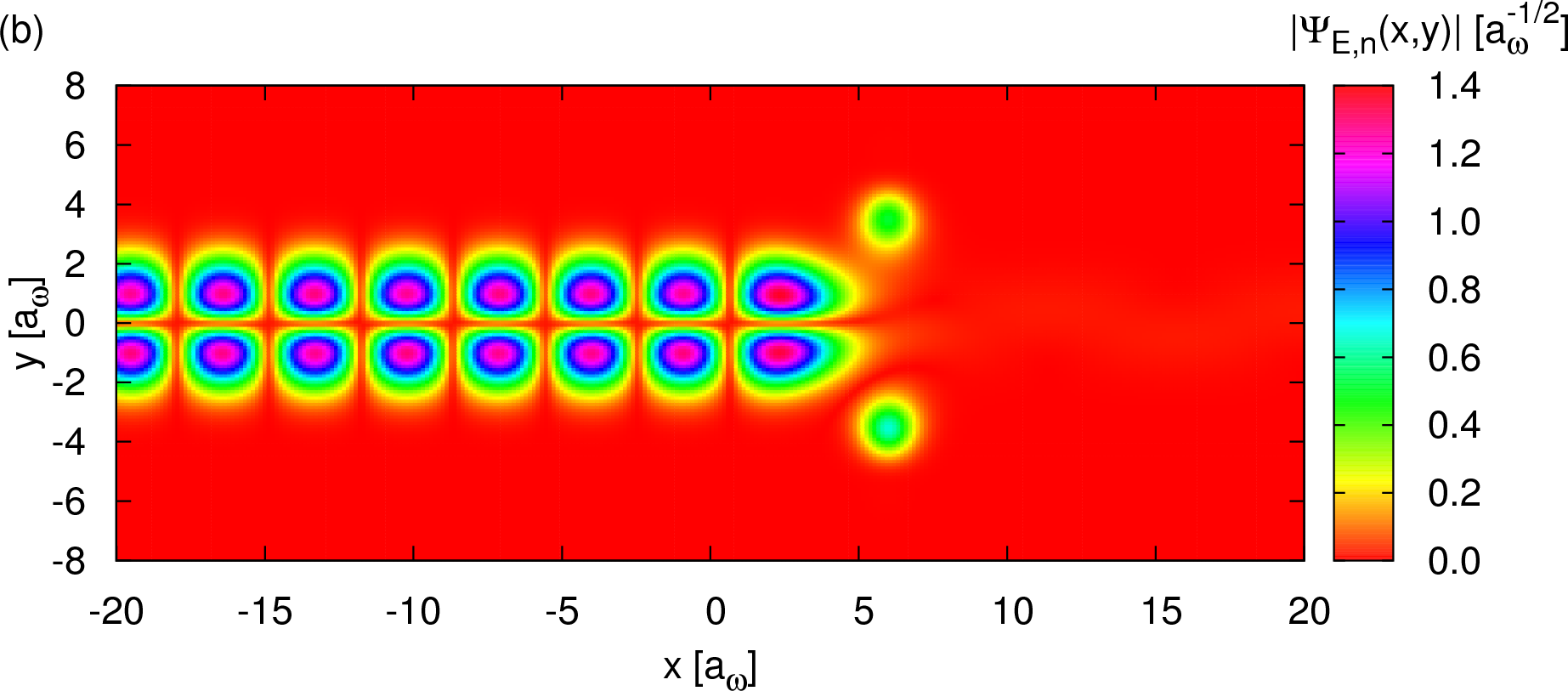}
 \caption{(Color online) Square root of the probability densities of the scattering states
          marked by \marked{f} in Fig.~\ref{Fig:TwoP_SCQD_EneData_DT2} for the case of $\Delta V_{2a} = -0.14 \,
          \varepsilon_0$ with incident subband (a) $n$=$0$; and (b) $n$=$1$.}
 \label{Fig:Psi_f}
\end{figure}

\begin{figure}[htbq]
 \centering
 \includegraphics[width=0.45 \textwidth]{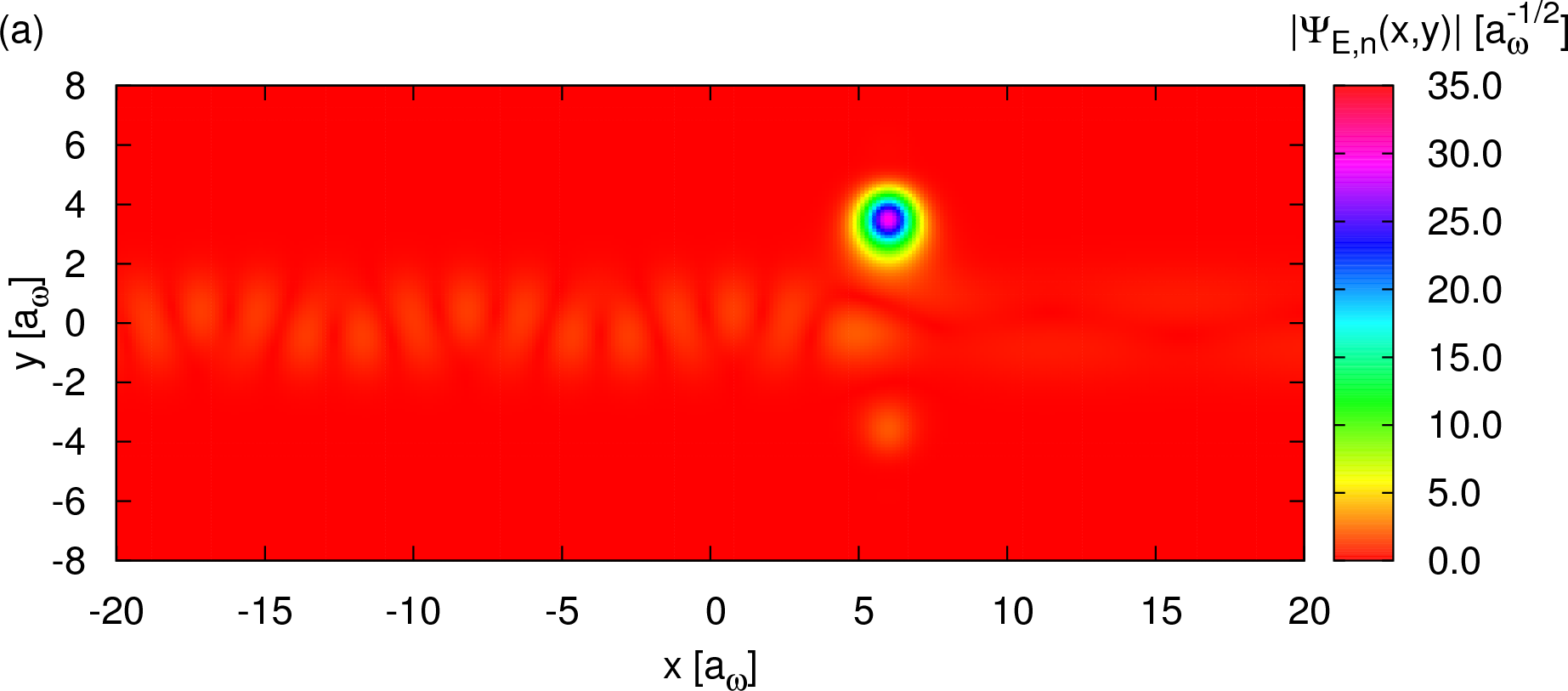}
 \includegraphics[width=0.45 \textwidth]{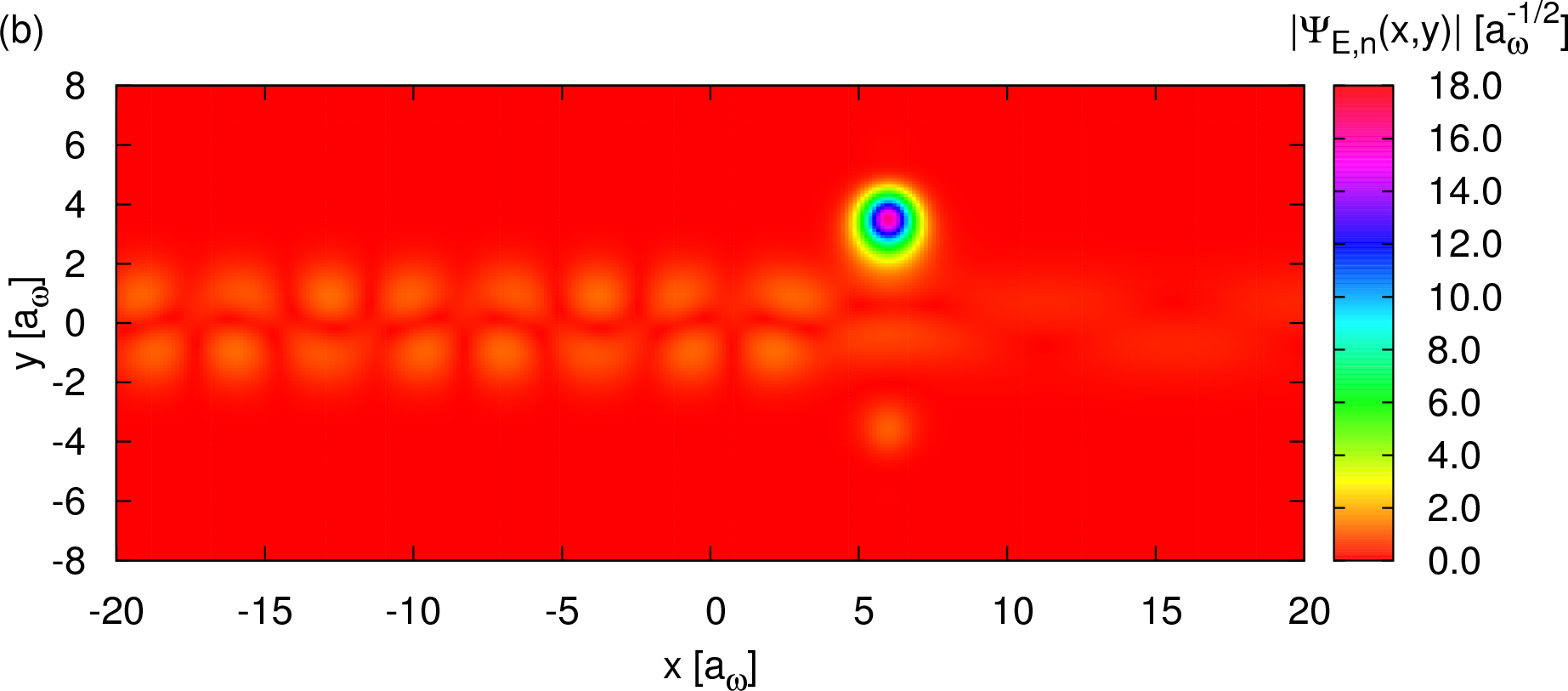}
 \caption{(Color online) Square root of the probability densities of the scattering states
          at the dip structure in conductance marked by \marked{g} in
          Fig.~\ref{Fig:TwoP_SCQD_EneData_DT2}
          for the case of $\Delta V_{2a} = -0.14 \,
          \varepsilon_0$ with incident subband (a) $n$=$0$; and (b) $n$=$1$.}
 \label{Fig:Psi_g}
\end{figure}

In Fig.\ \ref{Fig:TwoP_SCQD_EneData_DT2}, we present the conductance
characteristics with detuning parameter $\Delta V_{2a} = -0.14 \,
\varepsilon_0$ (solid red) in comparison to the case without
detuning $\Delta V_{2a} = 0.0 \, \varepsilon_0$ (dashed blue).
Furthermore, in Figs.\ \ref{Fig:Psi_e}, \ref{Fig:Psi_f}, and
\ref{Fig:Psi_g}, we present, in order of energy, the square root of
the probability densities of the scattering states at the energies
labeled by \marked{e}, \marked{f}, and \marked{g}, respectively, in
Fig.\ \ref{Fig:TwoP_SCQD_EneData_DT2}. In the presence of detuning
$\Delta V_{2a} = -0.14 \, \varepsilon_0$, two significant
Breit-Wigner dips are found at $E$ = $3.9725 \, \varepsilon_0$
(labeled by \marked{e}) and $4.0798 \, \varepsilon_0$ (labeled by
\marked{g}).  It turns out that the anti-resonant dip features are
significantly different from the antisymmetric Fano resonance for the
symmetric case in the absence of detuning.  It is worth mentioning
that these Breit-Wigner structures in conductance are associated to
the long-lived quasibound states in the lower dot for case
\marked{e} and in the upper dot in for case \marked{g}, as
illustrated in Fig.\ \ref{Fig:Psi_e} and Fig.\ \ref{Fig:Psi_g}
respectively. This important feature demonstrates the possibility of
\textit{coherent switching} between the lower and the upper
side-coupled dots by appropriately adjusting the electronic energy.
This could have interesting applications in quantum interference
devices based on coherent control of the quantum interference of
side-coupled double quantum dots.

We would like to bring attention in passing that the real space
probability density of the scattering states shown in Fig.\
\ref{Fig:Psi_f} corresponding to the
point \marked{f} of solid red curve in Fig.\
\ref{Fig:TwoP_SCQD_EneData_DT2} has significant coupling between the
extended propagating modes in the quantum channel and both the
side-coupled quantum dots.  In addition, the quasibound states
localized in each quantum dots are considerably suppressed and the
scattering behavior is much weaker than both the cases \marked{e}
and \marked{g}.

\begin{figure}[htbq]
 \centering
 \includegraphics[width=0.45 \textwidth]{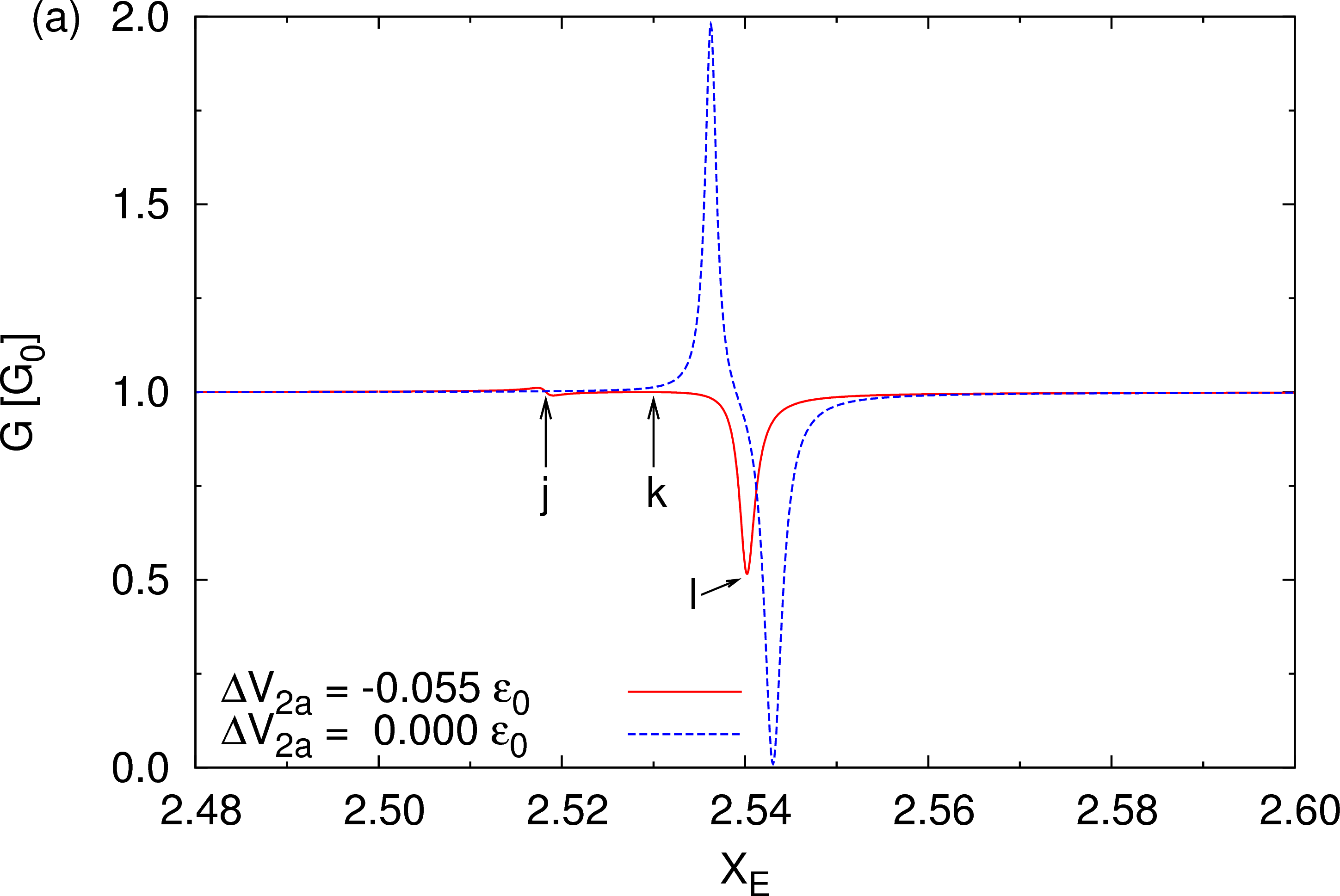}
 \includegraphics[width=0.45 \textwidth]{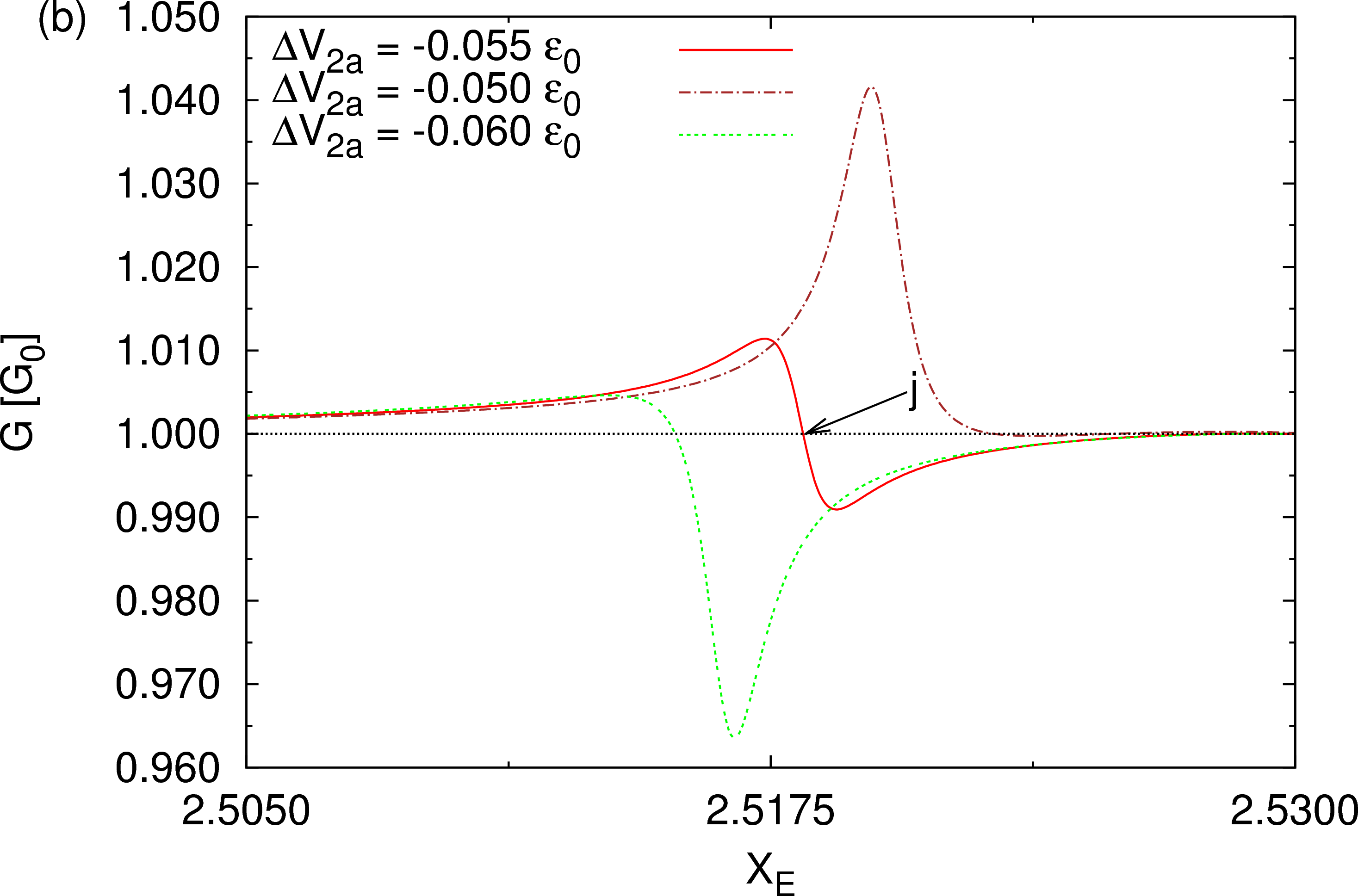}
 \caption{(Color online) (a) Conductance as a function of the
          parameter $X_{E}$ in a nanowire with double side-coupled
          quantum dots for the cases with detuning
          $\Delta V_{2a} = -0.055 \, \varepsilon_0$ (solid red) and
          with no detuning $\Delta V_{2a} = 0.0 \, \varepsilon_0$ (dashed
          blue). The small blip structure marked by \marked{j} in (a) is
          emphasized in (b)}
 \label{Fig:TwoP_SCQD_EneData_DT3}
\end{figure}
\begin{figure}[htbq]
 \centering
 \includegraphics[width=0.45 \textwidth]{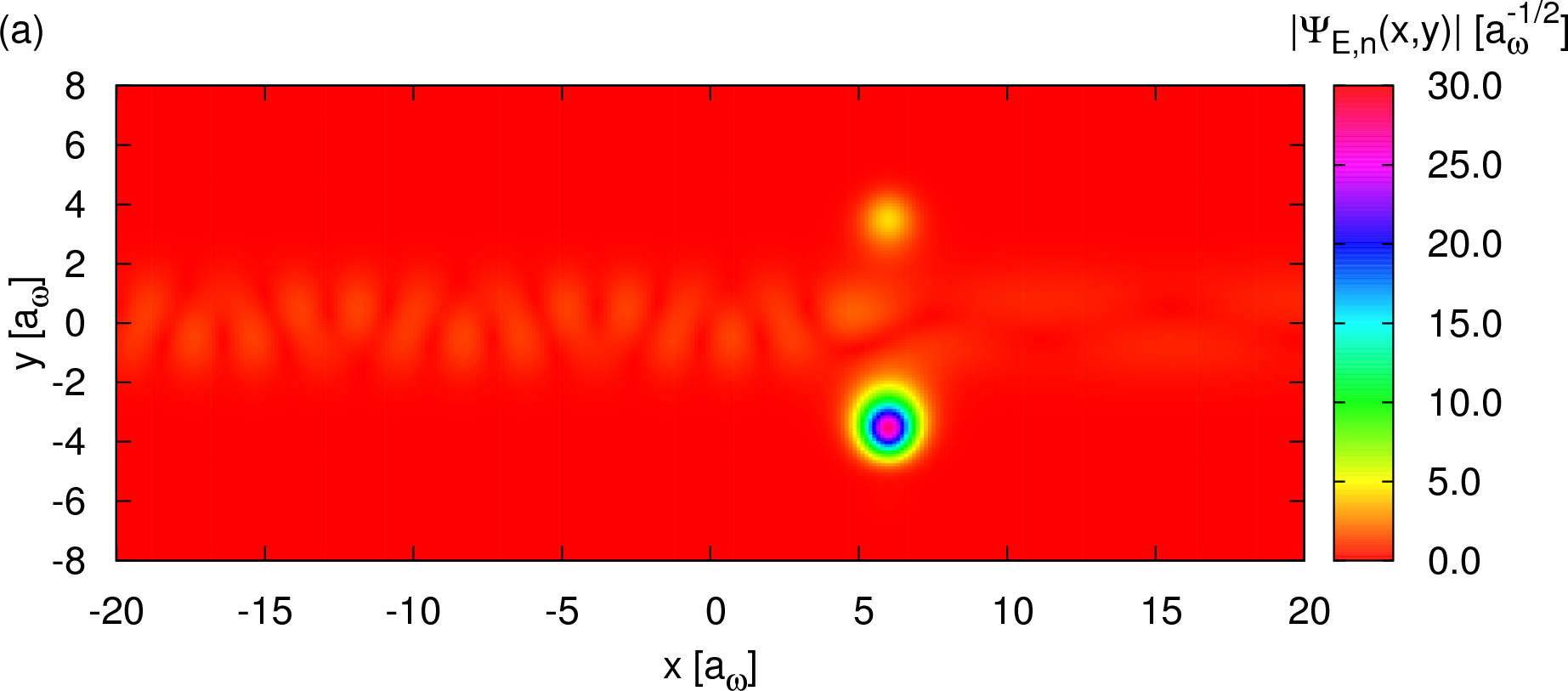}
 \includegraphics[width=0.45 \textwidth]{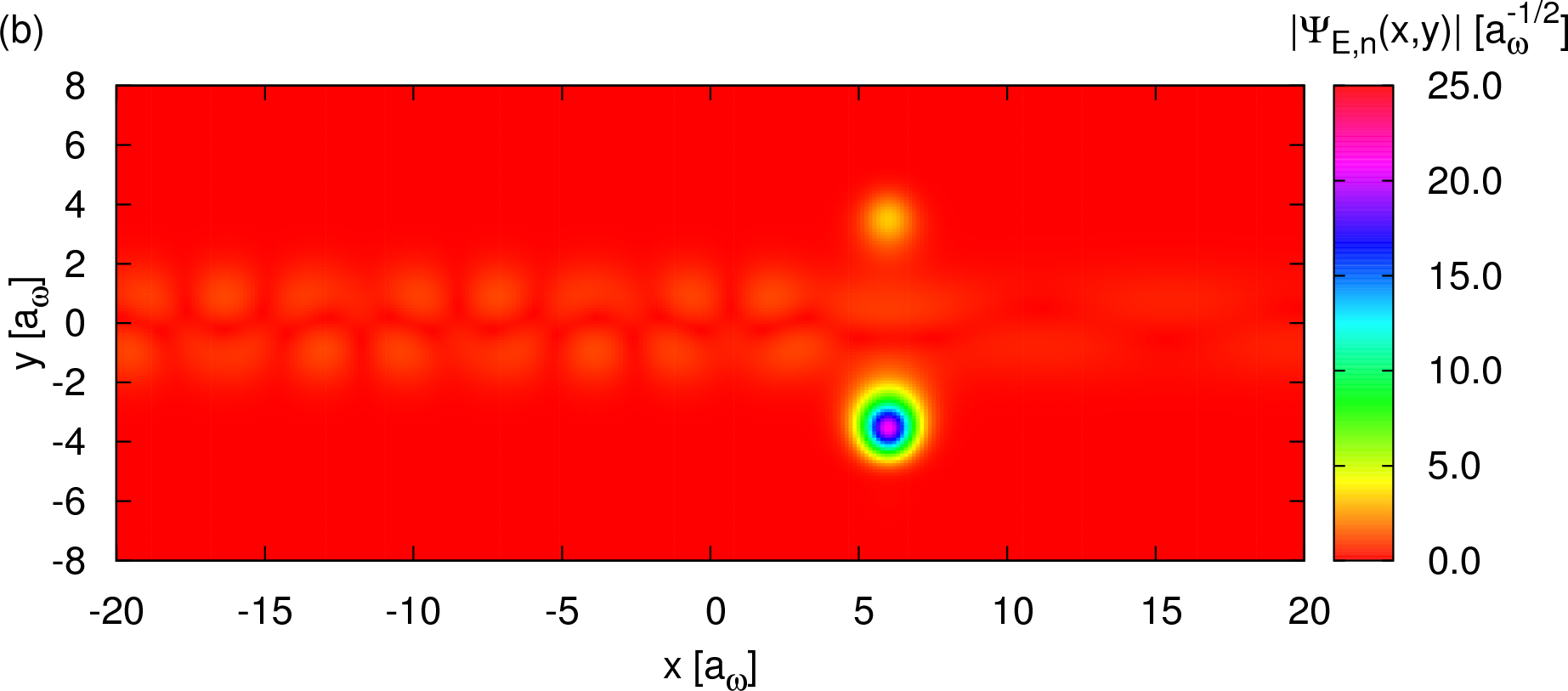}
 \caption{(Color online) Square root of the probability densities of the scattering states
          at the small blip in conductance marked by \marked{j} in
          Fig.~\ref{Fig:TwoP_SCQD_EneData_DT3}
          with detuning parameter $\Delta V_{2a} = -0.055 \, \varepsilon_0$
          for the cases of incident subband
          (a) $n$=$0$; and (b) $n$=$1$.}
 \label{Fig:Psi_j}
\end{figure}
\begin{figure}[htbq]
 \centering
 \includegraphics[width=0.45 \textwidth]{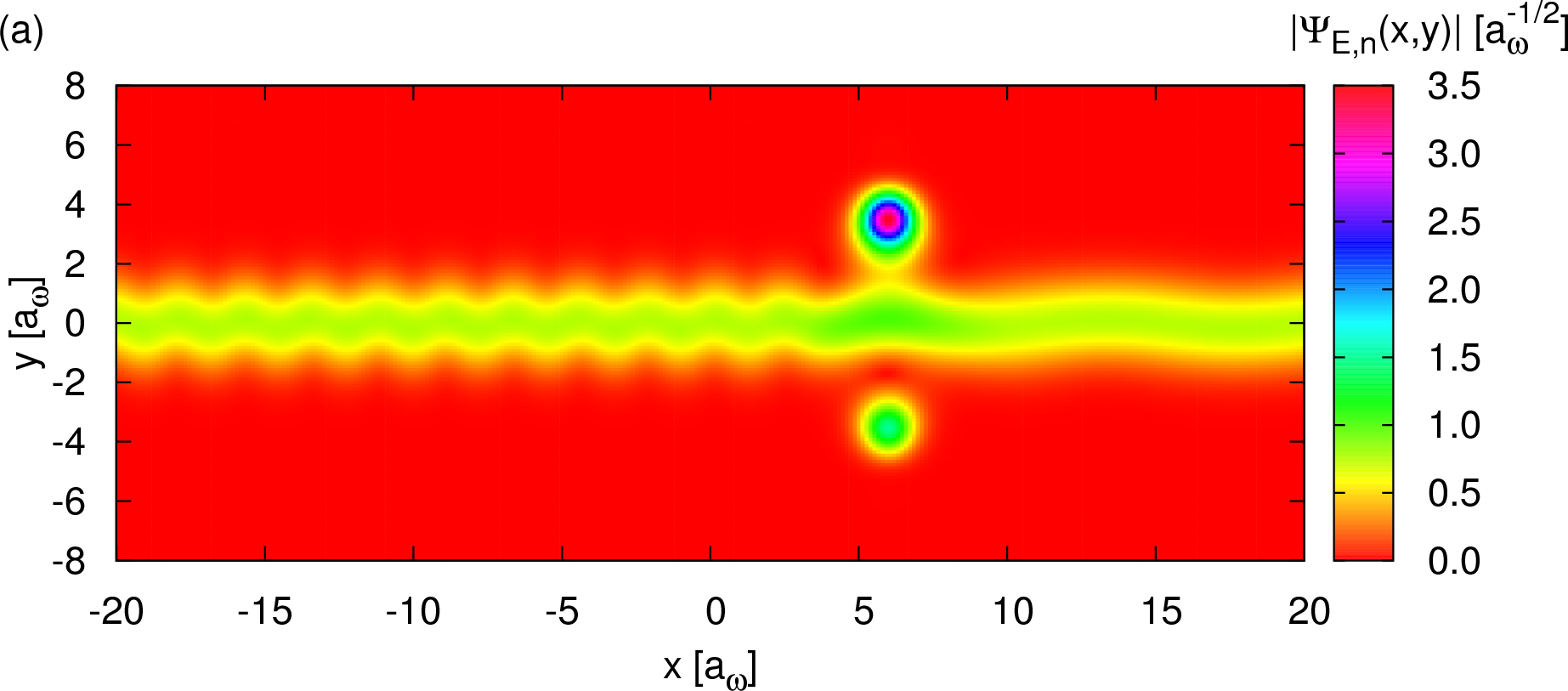}
 \includegraphics[width=0.45 \textwidth]{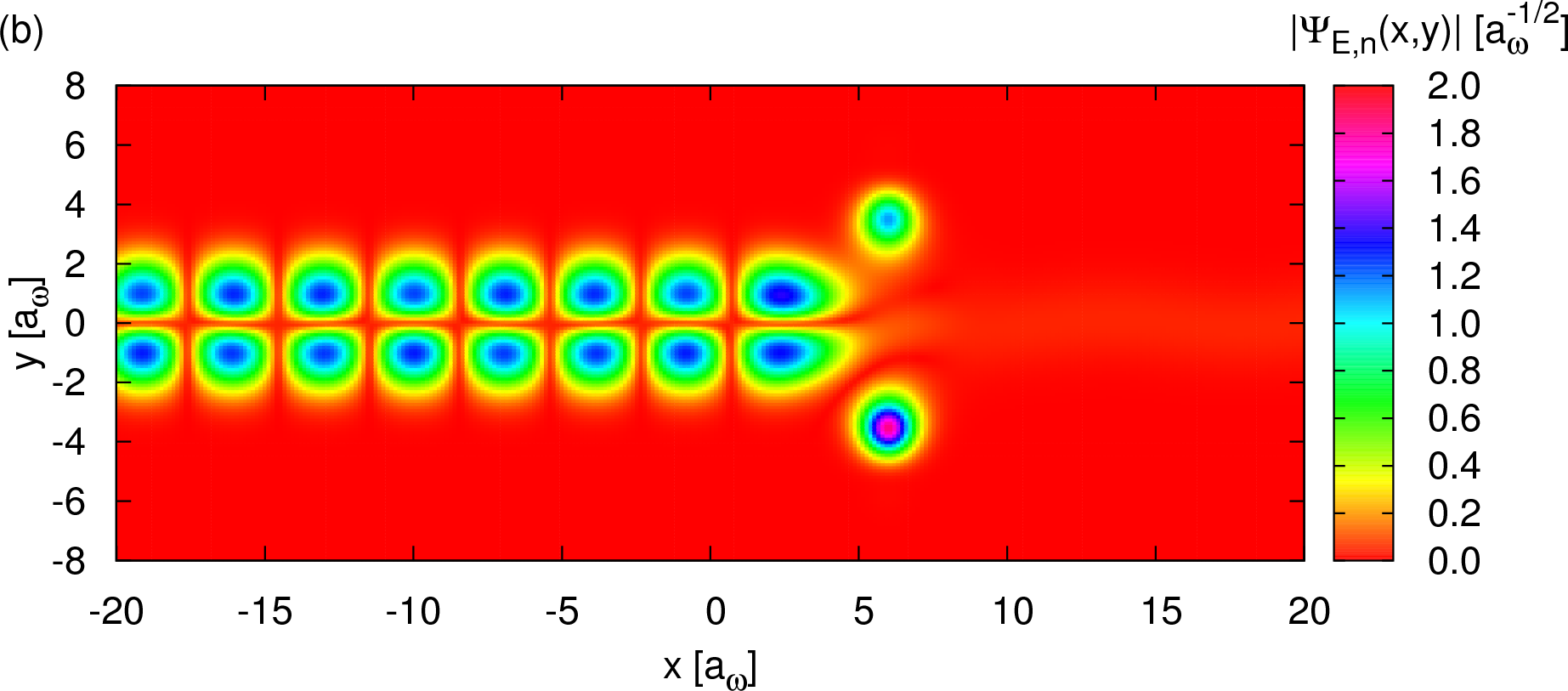}
 \caption{(Color online) Square root of the probability densities of the scattering states
          marked by \marked{k} in Fig.~\ref{Fig:TwoP_SCQD_EneData_DT3}
          with detuning parameter $\Delta V_{2a} = -0.055 \, \varepsilon_0$
          for the cases of incident subband
          (a) $n$=$0$; and (b) $n$=$1$.}
 \label{Fig:Psi_k}
\end{figure}

The peak-to-dip crossover feature demonstrated in Fig.\
\ref{Fig:TwoP_SCQD_EneData_DT1} motivates us to investigate further
the processes around  $\Delta V_{2a} \approx -0.05 \,
\varepsilon_0$ where only a single peak is significantly visible.
In Fig.\ \ref{Fig:TwoP_SCQD_EneData_DT3}(a), we thus focus on the
conductance as a function of the rescaled electron energy for the case
of $\Delta V_{2a} = -0.055 \, \varepsilon_0$ that is presented by
solid red curve in comparison with the case with no detuning shown
by dashed blue curve.  First, we find a small blip marked by
\marked{j} ($E=4.0365 \, \varepsilon_0$). A zoom-in figure for
clearly demonstrating the blip structure is illustrated in Fig.\
\ref{Fig:TwoP_SCQD_EneData_DT3}(b) in comparison with the cases of
$\Delta V_{2a}$ = $-0.050 \, \varepsilon_0$ (dash-dotted brown) and
$-0.060 \, \varepsilon_0$ (dotted green) for showing the sensitivity
nature of the blip structure in conductance.  To understand better
the transport mechanism of the blip structure marked by \marked{j},
it is helpful to compare with the square root probability density in
the real space shown in Fig.\ \ref{Fig:Psi_j}.  Electrons entering
the system in the lowest two subbands mainly form long-lived
quasibound states in the lower side-coupled quantum dot 1.  Second,
we would like to see the point marked by \marked{k} that is an
energy regime between the blip structure \marked{j} and the dip
structure \marked{l}. The localized electrons tend to be coupled to the
central quantum channel as shown in Fig.\ \ref{Fig:Psi_k}.  The
$n$=$0$ mode exhibits stronger coupling to the upper dot and
manifests a high transmission feature, however the $n$=$1$ mode with
higher coupling to the lower dot exhibits a total reflection.

Concerning the point marked by \marked{l} ($E=4.0804 \,
\varepsilon_0$), the electrons occupying the lowest two subbands
tend to form long-lived quasibound states in the upper dot.  These
interesting features accessible by adjusting the electronic energy
from \marked{j} to \marked{l} may be applicable for \textit{coherent
switching} of charge accumulation between the lower and the upper
dot.   We note that the small blip marked by \marked{j} can induce a
strong and long-lived bound state in quantum dot 2. This is not
straight forward expected from the relationship of lifetime and
full-width at half-maximum of the resonance structures in the
conductance. This counterintuitive feature in conductance
is related to an occurrence of a bound state in continuum (BIC)
in the double side-coupled quantum dot system.

BIC is a discrete energy normalizable bound state above
the continuum threshold in energy, having vanishing
resonance width and in principle infinite lifetime.
Shortly after that the Schr\"odinger equation was put forward were
von Neumann and Wigner the first to predict the existence of
the BIC phenomenon.\cite{Neumann1929} There, the BIC behavior was
regarded as a mathematical curiosity attributed to rather unphysical
spherically symmetric potentials. Later on, more detailed analysis
of the BIC was presented by Stillinger and Herrick.\cite{Stillinger1975}
Friedrich and Wintegen showed that the BIC may occur due
to the interference of resonances that can be analyzed using the
Feshbach resonance theory:\cite{Friedrich1985} Suppose two
resonances can be tuned as a function of a continuous parameter, for
a certain condition one of the resonances may have a vanishing
resonant width. This fact was discussed in a number of physical
systems both with analytical and numerical
models.\cite{Ordonez2006,Fan1998a,Guevara2003,Sadreev2006}
It is worth mentioning that the resonance width can also turn to
zero in open systems such as the angle of a bent
waveguide is varied.\cite{Olendski2002}
The BIC has also been discussed in the context of nuclear dynamics
on coupled potential surfaces.\cite{Cederbaum2003}
Furthermore, it has also been proposed that the BIC
causes a very narrow absorption peak in a semiconductor
heterostructure superlattice.\cite{Capasso1992}

In our system, the two resonant states are the quasibound states in the
two side-coupled quantum dots and the continuous parameter is the
detuning parameter $\Delta V_{2a}$.  However, it has not
been reported that a BIC structure in conductance can be controlled
by appropriately detuning one of a side-coupled quantum dots.
Furthermore, we would like to bring
attention to that the strong quasibound state at the blip
structure in the conductance is not the BIC but an indication of the
existence of such a state in the double side-coupled quantum
dot system, since the BIC or the ``ghost Fano resonance''
feature has been predicted to occur only when the configuration
is totally symmetric.\cite{Guevara2003}

\begin{figure}[htbq]
 \centering
 \includegraphics[width=0.45 \textwidth]{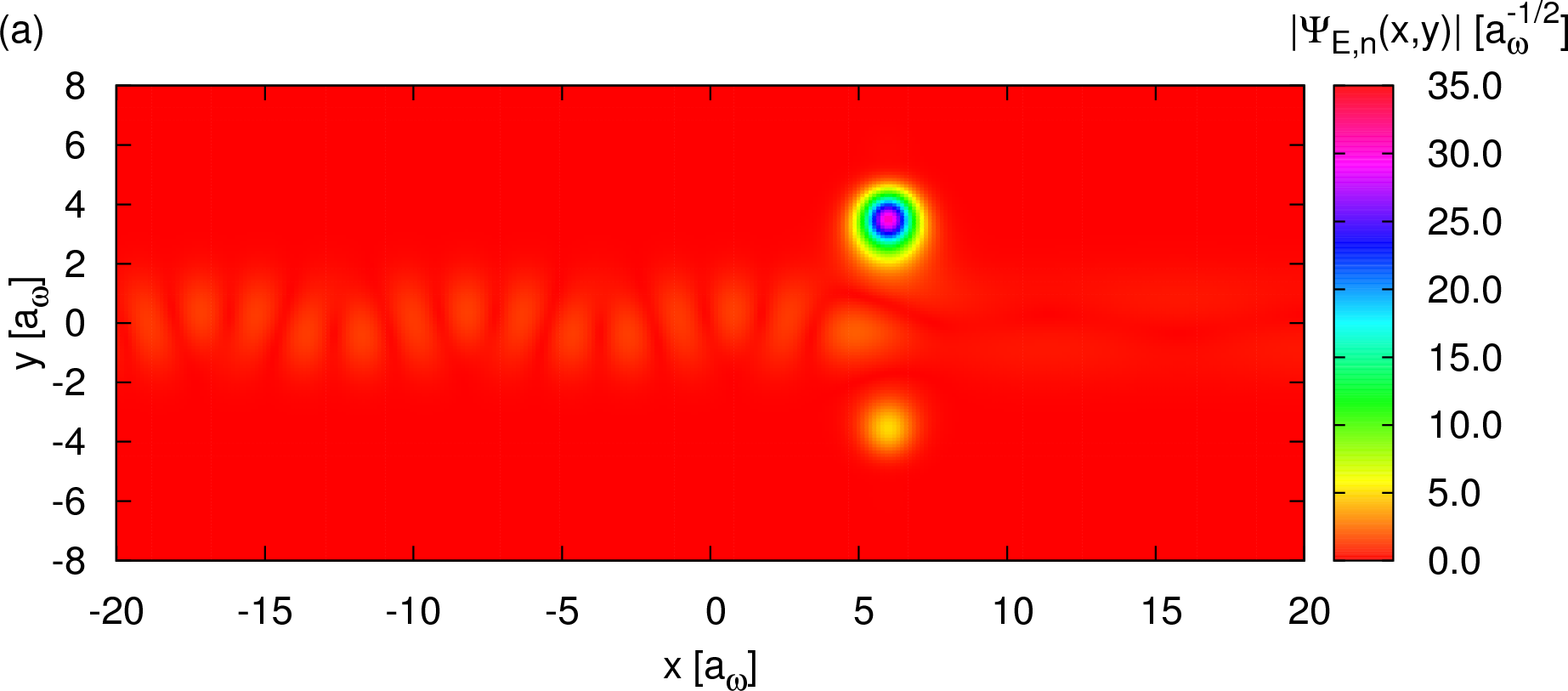}
 \includegraphics[width=0.45 \textwidth]{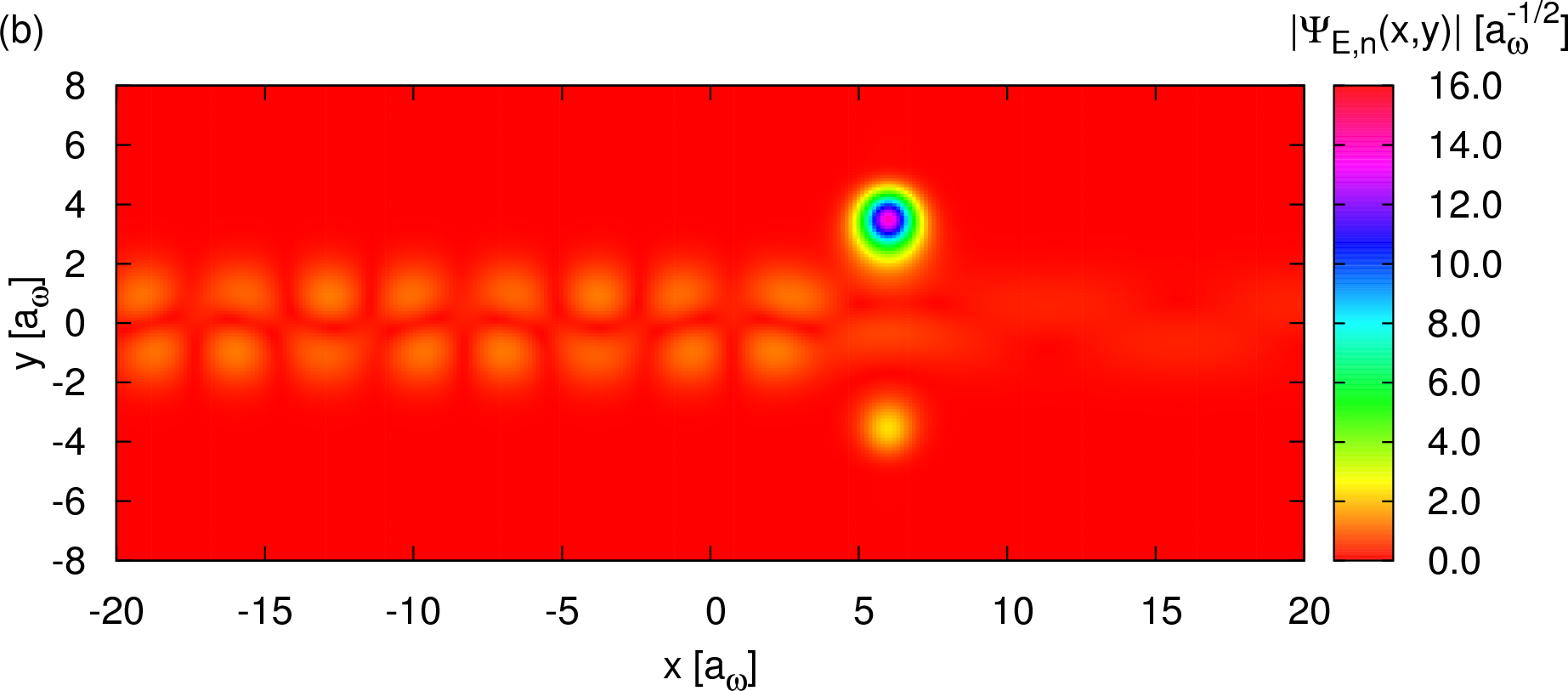}
 \caption{(Color online) Square root of the probability densities of the scattering states
          at the dip marked by \marked{l} in
          Fig.~\ref{Fig:TwoP_SCQD_EneData_DT3} with detuning parameter $\Delta V_{2a} = -0.055 \, \varepsilon_0$
          for the cases of incident subband (a) $n$=$0$; (b) $n$=$1$.}
 \label{Fig:Psi_l}
\end{figure}

\section{Concluding Remarks}
We have developed a theoretical model by implementing the
Lippmann-Schwinger formalism to demonstrate and elucidate the
quantum transport properties of a quantum dot systems side-coupled
to a nanowire.    In the present work, we propose that the
conductance can be switched on and off by turning on and off one of
the side-coupled double quantum dots controlled by using a gate
voltage.  It has been verified experimentally that a conductance
switching can be achieved either in a conducting polymer wire by
controlling the redox state of the polymer with a potential of the
nanoelectrodes.\cite{He2003} Theoretically, Taylor and co-workers
also investigated the conductance switching features in a molecular
device for studying the effect of side groups on the electrical
properties of a monolayer.\cite{Taylor2003}

Furthermore, we have demonstrated the possibility of using the
detuning parameter to achieve \textit{coherent switching} between
the lower and the upper side-coupled dots by appropriately
adjusting the electronic energy.  This may be utilized in designing
quantum devices based on the coherent control of the side-coupled
dots.  For the case of slight detuning, say $\Delta
V_{2a} = -0.055$, we have found a small blip structure in
conductance that is very sensitive to the detuning parameter.
Coherent switching of charge accumulation for this case is proposed
and its relation to the BIC feature is also discussed.  Similar
coherent switching features can be found in the case of quadrupole
side-coupled quantum-dot system by means of detuning technique, but
the analysis is more difficult due to the complicated quantum
interference among the four dots. The significantly different
conductance properties, however, will be discussed elsewhere.  In
order to reveal the coherence switching and the dynamics of
electronic transport in a side-coupled quantum-dot system, more
intricate measurements are required. We wish that the above proposed
switching effects will stimulate experimental efforts in the
near future.

\begin{acknowledgments}
The authors acknowledge financial support from
the Icelandic Research Fund for Graduate Students,
the Icelandic Research Fund,
the Research Fund of the University of Iceland
the Icelandic Instrument Fund,
the Icelandic Science and Technology Research Program
for Postgenomic Biomedicine, Nanoscience and
Nanoscience and Nanotechnology,
and the Taiwan National Science Council under
the grant number NSC97-2112-M-239-003-MY3.
\end{acknowledgments}



\end{document}